\documentclass[11pt,a4paper]{article}
\usepackage[a4paper, margin=3cm]{geometry}
\usepackage{import}
\usepackage{dsfont}
\usepackage{caption}

\usepackage{amssymb,amsmath,amsfonts}
\usepackage{xcolor}
\usepackage{natbib}
\usepackage{authblk}

\usepackage{graphicx}
\usepackage{hyperref}
\hypersetup{
    colorlinks=true,
    linkcolor= blue,
    citecolor = blue
    }

\usepackage{algorithm, algpseudocode, subcaption}

\usepackage{xargs}

\newcommand{\br}[1]{\hspace{-0.2em} \left( #1 \right)}
\newcommand{\set}[1]{\left\{ #1 \right\}}

\newcommand{\Var}[1]{\text{Var}\left( #1 \right)}

\renewcommand{\d}[1]{\ensuremath{\operatorname{d}\!{#1}}}

\newcommand{\E}[1]{\mathbb{E} \hspace{-0.1em} \left[ #1 \right]}
\newcommand{\Tr}{^{\mbox{\tiny T}}}

\newcommand{\bs}[1]{\boldsymbol{#1}}

\usepackage{setspace}

\newcommand{\cond}{\,|\,}

\begin{document}
\title{Gaussian approximations for fast Bayesian inference of partially observed branching processes with applications to epidemiology}

\author[1]{Angus Lewis}
\author[1]{Antonio Parrella}
\author[1]{John Maclean}
\author[1,*]{Andrew J. Black}
\affil[1]{School of Computer and Mathematical Sciences, The University of Adelaide, Adelaide, SA 5005.}

\maketitle

\onehalfspacing
\begin{abstract}
    We consider the problem of inference for the states and parameters of a continuous-time multitype branching process from partially observed time series data. 
   Exact inference for this class of models, typically using sequential Monte Carlo, can be computationally challenging when the populations that are being modelled grow exponentially or the time series is long.
    Instead, we derive a Gaussian approximation for the transition function of the process that leads to a Kalman filtering algorithm that runs in a time independent of the population sizes. We also develop a hybrid approach for when populations are smaller and the approximation is less applicable. We investigate the performance of our approximation and algorithms to both a simple and a complex epidemic model, finding good adherence to the true posterior distributions in both cases with large computational speed-ups in most cases. We also apply our method to a COVID-19 dataset with time dependent parameters where exact methods are intractable due to the population sizes involved.
    \medskip

\end{abstract}

\section{Introduction}
Branching processes are a flexible stochastic model of populations with many applications across a range of scientific disciplines. For example, epidemiology \citep{ panaretosPartiallyObservedBranching2007,allenApplicationsMultiTypeBranching2015}, ecology \citep{jagersBranchingProcessesBiological1975,hautphenne2019}, evolutionary biology \citep{kimmelBranchingProcessesBiology2002, xuEfficientTransitionProbability2015} and economics \citep{frenkenTheoreticalFrameworkEvolutionary2007}. They are a popular model due to their analytic tractability stemming from the fundamental assumption that the agents or particles that comprise a population evolve independently once created \citep{harrisTheoryBranchingProcesses1963,modeMultitypeBranchingProcesses1971,athreyaBranchingProcesses1972}. The assumption that population evolve independently once created can be justified for many systems, especially for initially small invading populations that are far away from the carrying capacity of the system. Such conditions often lead to exponentially growing populations that are a fundamental feature of branching processes \citep{kimmelBranchingProcessesBiology2002}. 
Exponential dynamics also mean that inference for the states and parameters of branching process models---given (partial) time-series data on the populations---can be challenging due to the amount of computation required, which also grows exponentially with the length of the time-series. 

Rather than being an impediment, exponential growth dynamics also offer an opportunity for applying asymptotic approximations. The main contribution of this paper (see Section \ref{SEC::Transition Density}) is to apply Gaussian approximations which enable the analytic and computationally tractable description of the evolution of the state of a continuous-time multitype branching process \citep{dormanGardenBranchingProcesses2004}. State inference can then be performed analytically using the well-known Kalman filter \citep{sarkkaBayesianFilteringSmoothing2013} and without the need for costly simulations or sampling that current methods utilise \citep{wilkinsonStochasticModellingSystems2018,sisson2019,cranmerFrontierSimulationbasedInference2020}. Moreover, a computationally efficient approximation to the marginal likelihood is also obtained which can be used within an Markov Chain Monte Carlo (MCMC) scheme to infer the states and parameters of the model. We demonstrate, via numerical experiments, that the resulting inference algorithm can be orders of magnitude faster than the current simulation-based approaches and that little bias in the posterior distributions is introduced by the approximation. 

Typically, as is the case in our application to epidemics, population sizes are initially small. Small populations can demonstrate more variable sample paths \citep{blackEffectClumpedPopulation2014,morris2023} and in these cases our approximation may not be justified. For these cases, we propose that a hybrid method can be employed. The hybrid method switches between a particle filter, which is used at times when the population size(s) is relatively small, and a Gaussian approximation, which is used at times when the population size(s) is relatively large. In Section \ref{SEC::Results} we demonstrate the accuracy and speed of our methods and apply them to epidemic models. We empirically investigate the biases introduced by the approximation. Code implementing the algorithms and examples is made freely available on github at \url{https://github.com/angus-lewis/MultitypeBranchingProcessInference.jl/tree/paper_examples}.

\section{Branching process model}
\label{sec:model}
Here we consider continuous-time multitype branching processes (CTBP) and to describe them we follow the notation of \cite{dormanGardenBranchingProcesses2004}. A CTBP models the evolution of a population of discrete \textit{agents}. A key assumption of a CTBP is that each agent in the population evolves independently of all others. In multitype branching processes, agents can be categorised as one of $r$ possible types. At any time, $t$, the state of the system is described by the random (row) vector $\mathbf{z}_t = (z_{1,t},\dots, z_{r,t}) \in (\mathbb{Z}^+)^r$, where $z_{i,t}$ is the number of agents of type $i$ at time $t$. 
Formally, a CTBP is the sequence \(\{\mathbf{z}_t\}_{t\geq0}\).
Once created, each agent lives for an exponentially distributed amount of time before dying and simultaneously giving birth to a random number of progeny or offspring. The rates of the lifetime distributions for each type are denoted $\boldsymbol{\omega} = (\omega_1, \omega_2, \dots , \omega_r)$. When an agent of type $i$ dies they produce $j_1, \dots, j_r$ agents of type $1,\dots, r$, respectively, with probability $p_{i,\,\mathbf{j}}$. Note, here $\mathbf{j} = (j_1, \dots, j_r)$ is a vector and is assumed to be fixed and known. 

Immigration can also be incorporated in CTBPs by modelling the arrival of new agents as a Poisson process. We denote the rate of immigration of agents of type $i$ as $\alpha_i$. Often, when an agent dies it also produces an agent of the same type; such an event can represent a birth without a corresponding death. Hence, CTBPs can capture a wide range of birth-death type dynamics. As the lifetimes are exponentially distributed, CTBPs are also continuous-time Markov chains (CTMC) with linear rates \citep{ross2000} and simulation of CTBP is straightforward using standard CTMC methods \citep{wilkinsonStochasticModellingSystems2018}. 

An important quantity in the analysis of CTBPs and the calculation of the the mean and variance of the process is the characteristic matrix $\boldsymbol{\Omega}$, with elements
\begin{equation}
    \label{EQN:: Omega Matrix}
    [\boldsymbol{\Omega}]_{ik}
    =
    \begin{cases}
        \omega_if_{ik}, & i \neq k, \\
        \omega_i(f_{ik} - 1), & i = k,
    \end{cases} 
\end{equation}
where $\displaystyle f_{ik} = \left[\sum_{\textbf{j}\in(\mathbb Z^+)^r}\textbf{j}\cdot p_{i,\textbf{j}}\right]_k$ 
is the expected number of agents of type $k$ produced on the event of a death of an agent of type $i$ \citep{dormanGardenBranchingProcesses2004}. It can be shown that the mean of the CTBP is given by \citep{dormanGardenBranchingProcesses2004}
\[E[\mathbf z_t\cond \mathbf z_0] = \mathbf{z}_0 e^{\bs \Omega t}\]
The dominant eigenvalue of $\boldsymbol{\Omega}$, denoted $\phi$, determines the long-term dynamics of the system \citep{modeMultitypeBranchingProcesses1971,athreyaBranchingProcesses1972}. At long times, the mean of the CTBP is approximately
$$
\mathrm{E} [\mathbf{z}_t] \propto  \mathbf{u} e^{\phi t} ,
$$
where $\mathbf{u}$ is the (left) eigenvector of $\boldsymbol{\Omega}$ corresponding to the eigenvalue $\phi$. If $\phi$ is positive then the numbers of agents grows exponentially (known as the super-critical regime) and the proportions of each type settle into a fixed ratio. The eigenvalue $\phi$ is often called the early growth rate in epidemiology, or the Malthusian parameter in ecology. A summary of the calculation of the mean and variance of the process, \(\{\mathbf z_t\}_{t\geq0}\), which are needed for later derivations, is given in \citet{dormanGardenBranchingProcesses2004} with specific details also covered in Appendix \ref{APP::Branching Process Mean and Variance Calculation}.

\subsubsection*{Epidemic example}
To illustrate and clarify the CTBP concepts so far, we give an example of a CTBP for modelling the early stages of an epidemic. In the next section we will continue with this example to demonstrate how observations of the system are modelled. 

Consider the early stages of an infectious disease epidemic spreading through a large susceptible population. Individuals can either be susceptible (they do not have the disease yet), exposed (they have the disease but are not yet infectious/symptomatic), infectious (have the disease and can spread it), or recovered/removed (unable to get the disease) \citep{allenApplicationsMultiTypeBranching2015}. Under these circumstances the proportion of susceptible individuals is approximately 1 and the proportion of recovered individuals is approximately 0. Hence, it is reasonable to model the only the exposed and infectious individuals and neglect the susceptible and recovered individuals. Suppose that, while they remain infectious, each infectious individual spreads the disease to susceptible individuals independently and at a constant rate \(\beta\). Individuals who are infected are initially classified as exposed and are asymptomatic. These individuals remain exposed for a random amount of time which is exponential with rate \(\delta\) before becoming symptomatic and being classified as infectious. Individuals remain infectious for a random amount of time which is exponential with rate \(\lambda\).

To model this system consider a CTBP model with two types, susceptible and infectious. The state vector of the CTBP is \(\textbf{z}_t=(E_t, I_t)\) where \(E_t\) and \(I_t\) are the numbers of exposed and infectious individuals at time \(t\), respectively. To capture the transition of exposed individuals to infectious set the rate at which exposed agents ``die'' to \(\omega_1=\delta\) and suppose that upon a death event, with probability 1, an infectious agent is created. The expected progeny upon the ``death'' of a exposed individual is \(\textbf{f}_1=(f_{1,1}, f_{1,2})=(0, 1)\). 

Let the rate at which infectious agents ``die'' be \(\omega_2=\beta+\lambda\). Upon the death of an infectious agent, with probability \(\beta/\omega_2\) an infection event happens and in this case a new exposed agent is created and the infectious agent replaces itself. Alternatively, upon the death of an infectious agent, with probability \(1-\beta/\omega_2=\lambda/\omega_2\) the infectious agent recovers, i.e., the infectious agent dies and no progeny are created. The expected number of progeny upon the ``death'' of an infectious agent is \(\textbf{f}_2=(f_{2,1}, f_{2,2})=(\beta/\omega_2, \beta/\omega_2)\).

The corresponding $\bs{\Omega}$ matrix is 
\begin{equation*}
    \boldsymbol{\Omega}
    =
    \begin{bmatrix}
        \omega_1 & 0 \\
        0 & \omega_2 
    \end{bmatrix}
    \left(\begin{bmatrix}
        0 & 1 \\
        \beta/\omega_2 & \beta/\omega_2 
    \end{bmatrix}
    - \begin{bmatrix}
        1 & 0 \\
        0 & 1 
    \end{bmatrix}\right)
    =
    \begin{bmatrix}
        -\omega_1 & \omega_1 \\
        \beta & -\lambda 
    \end{bmatrix}.
\end{equation*}

It can be shown that \(\bs \Omega\) has two distinct real eigenvalues and two linearly independent eigenvectors provided that \(\beta,\omega_1>0\) or \(\lambda>\omega_1\). In the case of two distinct eigenvalues, let \(r=\displaystyle\sqrt{4\beta\omega_1+(\lambda-\omega_1)^2}\), and \(\theta_1=\frac{1}{2}(-r-\lambda-\omega_1)\) and \(\theta_2=\frac{1}{2}(r-\lambda-\omega_1)\) be the eigenvalues of \(\bs \Omega\), and define \(\hat \theta_i=\lambda + \theta_i,\, i=1,2\), then it can be shown that the mean state, given an initial condition, is
\[E[\bs z_t\cond \bs z_0] = \bs z_0 \cfrac{1}{\hat \theta_1-\hat \theta_2}\begin{bmatrix}
    \hat\theta_1e^{\theta_1t} - \hat \theta_2e^{\theta_2t} & \cfrac{\hat\theta_1\hat\theta_2}{\beta}\left(e^{\theta_2t}-e^{\theta_1t}\right)\\
    \beta\left(e^{\theta_1t}-e^{\theta_2t}\right) & \hat\theta_1e^{\theta_2t}-\hat \theta_2e^{\theta_1t}
\end{bmatrix}.\] 
Further, if \(\theta_2\) is the maximal eigenvalue (the Malthusian parameter, \(\phi\)), then the eigenvector corresponding to \(\theta_2\) is proportional to 
\[
\begin{bmatrix}
    \beta \\ \omega_1 + \theta_2
\end{bmatrix}.
\]

\begin{figure}
\centering
    \includegraphics[width = \textwidth]{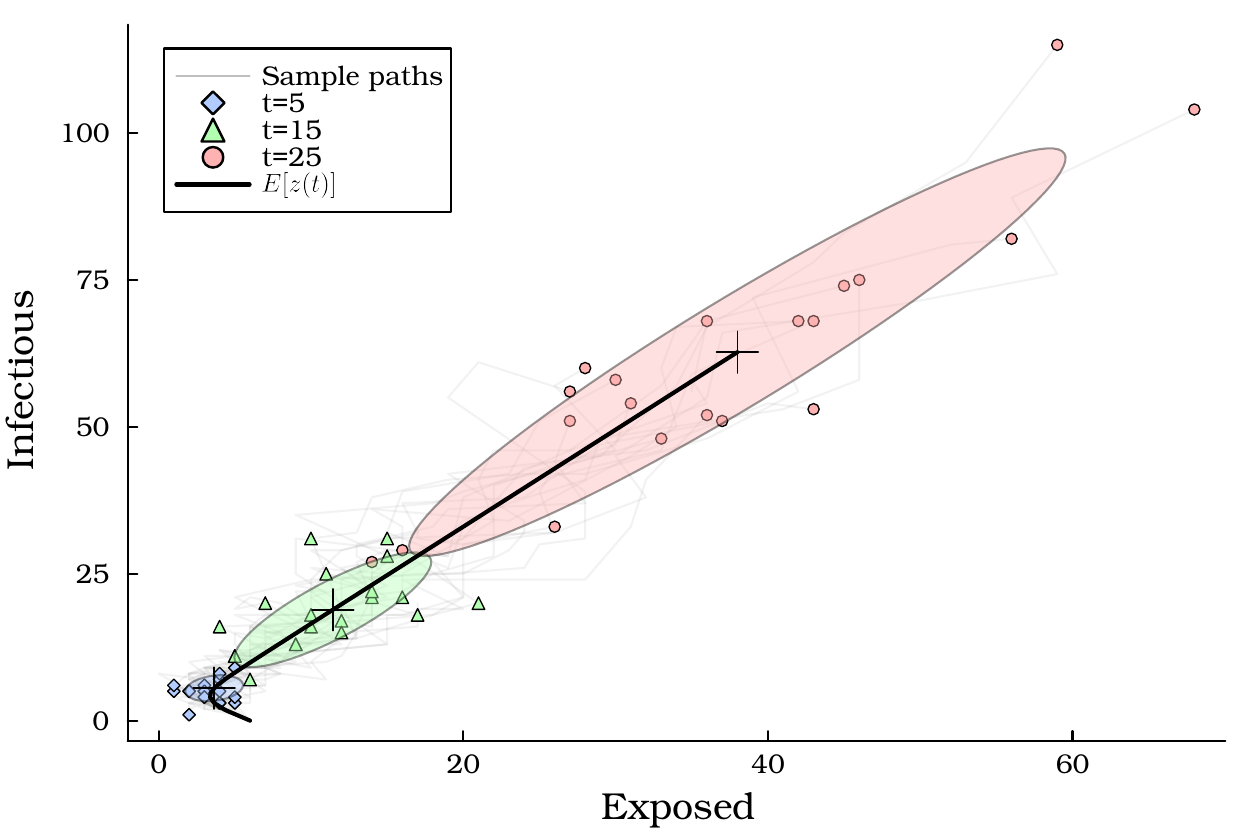}
    \caption{Simple SEIR epidemic model dynamics for exposed and infectious individuals.
    The grey lines show the full realisations, with the dots indicating the state of the realisations at $t = 5,\,15,\,25$. The solid black line is the true mean of the process for all time points. The crosses are the true means for the state of the process at times $t = 5,\,15,\,25$. The ellipses are the contours of the Gaussian approximations for the covariance evaluated at one standard deviation from the mean, centred at the crosses with covariance equal to that of the CTBP at the respective times.}
    \label{FIG::Branching Process Example}
\end{figure}

Figure~\ref{FIG::Branching Process Example} illustrates the simulated stochastic dynamics of this model along with analytically calculated values for the mean and variance (see Appendix \ref{APP::Branching Process Mean and Variance Calculation} for details of calculation) with parameters \(\beta=0.3\), \(\omega_1=\delta=0.375\) and $\lambda=3/28$, $R_0=2.8$, an initial population of 6 exposed individuals and 0 infectious individuals. The eigenvalues for the characteristic matrix of this model are \(\theta_1=-\sqrt{\frac{8181}{62720}}-\frac{27}{112}\approx -0.6022\) and \(\theta_2=\sqrt{\frac{8181}{62720}}-\frac{27}{112}\approx 0.1201\). The Malthusian parameter is \(\theta_2\) with a corresponding eigenvector given by
\[
\bs v_2 = \begin{bmatrix}
    \beta \\ \omega_1 + \theta_2
\end{bmatrix} = 
\begin{bmatrix}
    \frac{3}{10} \\ \sqrt{\frac{8181}{62720}} + \frac{15}{112}
\end{bmatrix}\approx 
\begin{bmatrix}
    0.3 \\ 0.495
\end{bmatrix} \propto 
\begin{bmatrix}
    1 \\ 1.65
\end{bmatrix}.
\]
For large \(t\) the tangent of the path \(E[\bs z_t]\) points approximately in the direction \(\bs v_2\), and \(E[\bs z_t]\) grows proportional \(e^{\theta_2t}\) in the direction \(\bs v_2\).

\subsection{Observation model}
In many applications the population counts of each agent type will not be directly observable. Here we introduce an observation model which describes how the observed data can be generated on top of, or as part of, the underlying CTBP model. 

Suppose that the state of the CTBP, $\set{\textbf{z}_t}_{t\geq 0}$ is not directly observed (i.e., is \textit{hidden}), and that the observed data is generated via the \textit{observation model}
\begin{equation}
    \mathbf{y}_t\cond\textbf{z}_t \sim \mathcal{N}(\mathbf{H}\textbf{z}_t\Tr, \mathbf{R}),
    \label{EQN::Obsevation dist requirement}
\end{equation}
where \({}\Tr\) denotes the transpose, $\mathbf{y}_t$ is a column vector of size $d$, $\mathbf{H}$ is a $d\times r$ matrix and $\mathbf{R}$ is a $d\times d$ covariance matrix. The matrix $\mathbf{H}$ maps the hidden state of the CTBP to the observation space and the covariance matrix $\mathbf{R}$ specifies the covariance of any additional noise on top of this. In this paper we assume that observations are made at fixed time steps which, without loss of generality, we assume is one unit. Hence we write our data as $\textbf{y}_{1:T} := \set{\textbf{y}_s}_{s=1}^T$. The methods here can be easily extended to non-uniform time intervals at the expense of extra computational time. 

The noise variance matrix, $\mathbf{R}$, can be interpreted as additional noise on top of the CTBP that captures physical characteristics of the sampling process used to observe the data. Alternatively, if observation noise is not a realistic assumption, then the additional noise of the observation model on top of the CTBP can be interpreted as a type of regularisation \citep{Musso2001,chopinIntroductionSequentialMonte2020} against model misspecification at the cost of some accuracy of the parameter posteriors.

For applications such as epidemic modelling, population numbers (e.g., the infectious prevalence) cannot be observed directly and instead only incidence data---the number of new cases per unit time---is actually observed \citep{fontain2019}. To generate incidence data the underlying CTBP can be augmented with a new type that counts the cumulative number of certain events. By setting the lifetime rate of this type to zero the value of the type is non-decreasing as a function of time. The number of new events over a time-step can then be calculated from the difference in the counting variables over the time-step. State-dependent noise can also be incorporated into the counting type by introducing a probability \(p\) with which events are observed. For example, with reference to the SEIR example above, we suppose that upon the birth of a new infectious individual the event is observed with probability \(p\) and the counting variable increases, and with probability \(1-p\) the event is unobserved and the counting variable remains constant.

Further complexities can be added to the observation model by including time-dependent transformations, \(\mathbf H_t\), or observation covariance \(\mathbf R_t\). The parameters of the observation model may even be something the practitioner wishes to estimate. Other observation characteristics can be incorporated into the CTBP directly. For example, delays in observing events can be included by introducing ``delay'' states in the CTBP \citep{shearer:2022}.







\subsection*{SEIR Example continued}
Here we continue our example from the previous section by specifying the observation model for our SEIR model. Assume that the event that an exposed individuals become infectious is observed with probability $p$. The CTBP model is augmented with an additional counter type, denoted $z_{3,t}$. The rate of the lifetime distribution of the counter types is zero and hence the counter type is a non-decreasing function of time. There are no ``death'' events for type 3 individuals, and therefore no progeny, so the progeny distribution for type 3 is arbitrary. For the sake of notation, we specify \(\bs f_3 = (0,0,1)\). For exposed individuals, upon a death event, with probability $1-p$ an unobserved infectious agent only is created, and with probability $p$ an observed infectious agent and agent of type 3 are created, thereby incrementing the counting state \(z_{3,t}\) by 1. The expected progeny upon the ``death'' of a exposed individual is now \(\textbf{f}_1=(f_{1,1}, f_{1,2}, f_{1,3})=(0, 1, p)\). No type 3 individuals are created upon a death event for an infectious agent; the expected number of progeny upon the ``death'' of an infectious agent is \(\textbf{f}_2=(f_{2,1}, f_{2,2}, f_{2,3})=(\beta/\omega_2, \beta/\omega_2, 0)\).

The characteristic $\bs{\Omega}$ matrix is now
\begin{equation*}
    \boldsymbol{\Omega}
    =
    \begin{bmatrix}
        \omega_1 & 0 & 0\\
        0 & \omega_2 & 0\\
        0 & 0 & 0
    \end{bmatrix}
    \left(\begin{bmatrix}
        0 & 1 & p\\
        \beta/\omega_2 & \beta/\omega_2 & 0\\
        0 & 0 & 1
    \end{bmatrix}
    - \begin{bmatrix}
        1 & 0 & 0\\
        0 & 1 & 0\\
        0 & 0 & 1
    \end{bmatrix}\right)
    =
    \begin{bmatrix}
        -\omega_1 & \omega_1 & p\omega_1\\
        \beta & -\lambda & 0\\
        0 & 0 & 0
    \end{bmatrix}.
\end{equation*}
The mean function can be calculated analytically for this model, as can the covariance matrix.

\begin{figure}
\centering
    \includegraphics[width = 0.9\textwidth]{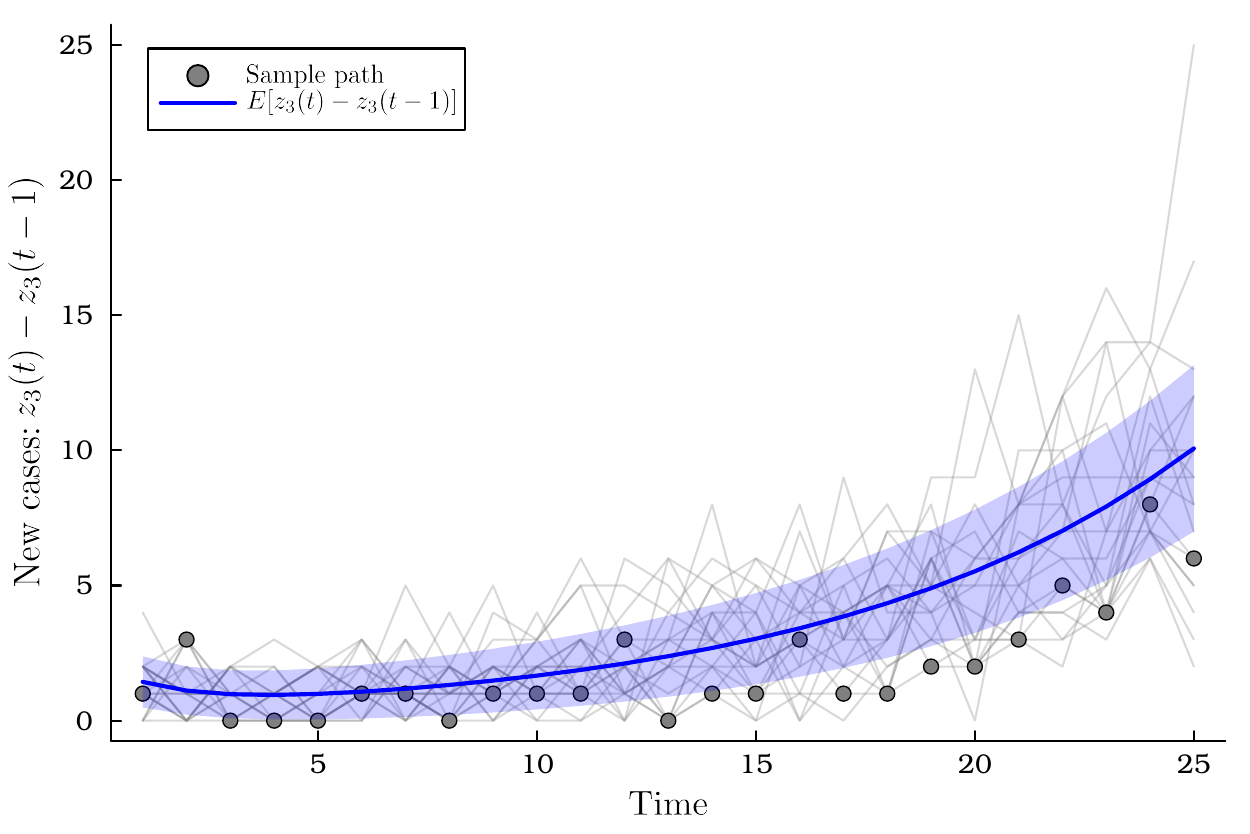}
    \caption{Simple SEIR epidemic model dynamics for new daily cases.
    The grey lines show sample paths of new daily cases for 20 realisations, with the dots highlighting a single realisation. The solid blue line is the true mean of new daily cases and the ribbons mark plus and minus one standard deviation from the mean.}
    \label{FIG::Branching Process Example Observed Cases}
\end{figure}

The process \(\{z_{t,3}\}_t\) is the cumulative number of observed infections by time \(t\). However, for our application to epidemic modelling, incidence data, the number of new observed infections each time period, is recorded. That is, we have observations of $z_{t,3}-z_{t-1,3}$. To get the new number of cases over a time step we set $z_{3,t}$ to zero at the beginning of each time step. This is allowed since the distribution of $z_{t,j},\,j\neq i$ and $z_{t,3}-z_{t-1,3}$, does not depend on $z_{t-1,3}$, given $z_{t-1,j},\, j\neq 3$.\footnote{Alternatively, by including an additional dependence on the state of the CTBP at the previous time step in the observation model we can capture new daily cases in the observation model instead of the CTBP, and the CTBP models the total cumulative number of cases. However, notation is simplified slightly if we reset the state $z_{t,3}$ at the beginning of each time-step. It is also possible to transform the daily case count observations to cumulative cases and model those instead, which requires mapping the observation model for daily case counts to the equivalent model for cumulative case counts, but it is more natural to model daily cases directly.} Figure~\ref{FIG::Branching Process Example Observed Cases} illustrates the stochastic dynamics of the observed number of new cases per time step along with analytically calculated values for the mean for the model with the same parameters as in Figure~\ref{FIG::Branching Process Example} with an observation state where cases are observed with probability \(p=0.75\).

The observation model for this process is specified by parameters \(\bs H = \begin{bmatrix}0&0&1\end{bmatrix}\) and \(\bs R=\begin{bmatrix}\sigma^2\end{bmatrix}\). Hence, the observations are \(\bs y_t = \begin{bmatrix}y_{t,1}\end{bmatrix} = (z_{t,3}-z_{t-1,3}) + \sigma \varepsilon_t\) where \(\varepsilon_t\sim N(0,1),\) independently.



\section{Inference Problem}
\label{sec:inference_problem}
A CTBP with an observation model as introduced in the previous sections can be formulated as a state-space model. Here we follow the setup and notation of \citet{sarkkaBayesianFilteringSmoothing2013}. Let \(\bs \theta\) denote the parameters of the model, then 
\begin{subequations}\label{EQN::HMM}
    \begin{align}
    \boldsymbol{\theta} &\sim p(\boldsymbol{\theta}),\\
    \textbf{z}_0 &\sim p(\textbf{z}_0), \\
    \textbf{z}_t &\sim p(\textbf{z}_t\cond \textbf{z}_{t-1},\boldsymbol{\theta}), \label{EQN::Trans kernal}\\
    \textbf{y}_t &\sim p(\textbf{y}_t \cond \textbf{z}_t, \boldsymbol{\theta}),
\end{align}
\end{subequations}
for $t \in \set{1,2,\dots, T}$. 
The inference problem is to sample from the joint posterior for the unknown states and parameters, $p(\mathbf{z}_{1:T}, \boldsymbol{\theta} \cond \mathbf{y}_{1:T})$. 
As we have a Markov process for the underlying model it is natural to use sequential filtering to evaluate the posterior distributions of the states of the hidden CTBP process, \(p(\mathbf z_t\cond \mathbf y_{1:t},\bs \theta)\). The marginal likelihood $p(\textbf{y}_{1:T} \cond \boldsymbol{\theta})$ is simultaneously calculated in such a procedure and we use the marginal likelihood inside a Metropolis Hastings (MH) scheme to sample the posterior of the parameters \citep{sarkkaBayesianFilteringSmoothing2013}.

Each step of the filtering process has three main steps. Starting with the posterior distribution of the state from the previous time step, \(p(\textbf{z}_t\cond \textbf{y}_{1:t},\boldsymbol{\theta})\) (or the initial distribution of the state, \(p(\textbf{z}_0\cond \boldsymbol{\theta})\) if at the first time-step), the distribution of the states at the next time step (state prediction) is calculated using 
\begin{equation}
    p(\textbf{z}_{t + 1}\cond \textbf{y}_{1:t}, \, \boldsymbol{\theta}) 
    = 
    \int_{\mathcal{S}} p(\textbf{z}_{t + 1}\cond \textbf{z}_t, \,\boldsymbol{\theta}) p(\textbf{z}_t\cond \textbf{y}_{1:t}, \,\boldsymbol{\theta}) \d{\textbf{z}_t}.
    \label{EQN::Prediction step}
\end{equation}
Next, using the observation model, \(p(\textbf{y}_{t + 1}\cond \textbf{z}_{t + 1}, \,\boldsymbol{\theta})\), and the state prediction distribution from \eqref{EQN::Prediction step}, the likelihood of the observation of $\textbf{y}_{t+1}$ (observation prediction) is calculated 
\begin{equation}
    p(\textbf{y}_{t+1} \cond \textbf{y}_{1:t}, \,\boldsymbol{\theta})
    =
    \int_{\mathcal{S}}
    p(\textbf{y}_{t + 1}\cond \textbf{z}_{t + 1}, \,\boldsymbol{\theta}) 
    p(\textbf{z}_{t + 1}\cond \textbf{y}_{1:t}, \,\boldsymbol{\theta}) \d{\textbf{z}_{t + 1}} 
    \label{EQN::Likihood step}.
\end{equation}
Finally, the filtering distribution (the posterior distribution of the state at time \(t+1\) given the observations up to and including time $t+1$) is then
\begin{eqnarray}
    p(\textbf{z}_{t + 1}\cond \textbf{y}_{1:t + 1}, \,\boldsymbol{\theta}) 
    &=& 
    \frac{p(\textbf{y}_{t + 1}\cond \textbf{z}_{t + 1}, \,\boldsymbol{\theta}) 
    p(\textbf{z}_{t + 1}\cond \textbf{y}_{1:t}, \,\boldsymbol{\theta})}{p(\textbf{y}_{t+1}\cond \textbf{y}_{1:t}, \,\boldsymbol{\theta})}.
    \label{EQN::Filtering step}
\end{eqnarray}
Equations \eqref{EQN::Prediction step}, \eqref{EQN::Likihood step} and \eqref{EQN::Filtering step} can be applied inductively for $t\in \set{1,2, \dots , T - 1}$. Using \eqref{EQN::Likihood step} for \(t=1,...,T,\) the marginal likelihood of the observations can be calculated as follows
\begin{eqnarray}
    p(\textbf{y}_{1:T} \cond \boldsymbol{\theta}) &=& 
    p(\textbf{y}_1 \cond \boldsymbol{\theta})\prod_{i = 1}^{T - 1}p(\textbf{y}_{t+1}\cond \textbf{y}_{1:t}, \,\boldsymbol{\theta}).
    \label{EQN::Full like calc}
\end{eqnarray}

The challenge in performing filtering for CTBP models is that the transition function, $p(\textbf{z}_{t+1} \cond \textbf{z}_{t},\boldsymbol{\theta})$, is generally intractable. As the state space of the branching process is countably infinite, it is possible to approximate the transition function by truncating the state space to a finite set, which then allows calculation via a matrix exponential \citep{blackCharacterisingPandemicSeverity2017,sherlockDirectStatisticalInference2021}. Of course, this works only when the populations are small, otherwise the size of the truncation space must grow with the population size (i.e., exponentially) at each time step, and the calculation again becomes intractable. A natural way around this is to instead use a sequential Monte Carlo approach for the filtering \citep{chopinIntroductionSequentialMonte2020}. 

In this paper, a basic bootstrap particle filter \citep{gordonNovelApproachNonlinear1993,chopinIntroductionSequentialMonte2020} is used to implement the Monte Carlo filtering. A bootstrap particle filter uses sequential-importance-resampling to propagate a set of weighted samples (of the state of the CTBP) that approximates the sequence of filtering distributions. Since the importance sampling distribution for each step is the transition density for the model, then the sampling is straightforward.  \cite{andrieuParticleMarkovChain2010} has shown that the likelihood estimates, $\hat{p}(\textbf{y}_{1:T}\cond \boldsymbol{\theta})$, produced by such a particle filter are unbiased and therefore the likelihood estimates can be used in place of the true likelihood within a Metropolis Hastings (MH) scheme to sample the posterior distribution of the parameter. This procedure is known as particle-marginal Metropolis Hastings (PMMH). 

As is the case with using the transition function for filtering, the bootstrap particle filter method also runs into computational problems when the observation sequence is long and/or when the population is large. For the MH chain to mix sufficiently, the variance of the likelihood estimate needs to be tightly controlled \citep{pittPropertiesMarkovChain2012,doucetEfficientImplementationMarkov2015,sherlockEfficiencyPseudomarginalRandom2015} which requires an increasing number of particles as the observation time series gets longer, or when the CTBP model is misspecified (as is almost always the case in practice). This is compounded when the CTBP is supercritical, so agent counts increase exponentially, and hence simulation costs also become exponentially more expensive \citep{golightlySimulationStochasticKinetic2013}. So although PMMH provides an exact approach to our inference problem, and hence a benchmark to test against, it still scales poorly to long observation sequences and large populations. 

\section{Approximation of the filtering recursions}
\label{SEC::Transition Density}
Given the challenges with evaluating the transition function described above, we next introduce a new Gaussian approximation to the transition function. This enables the filtering recursion calculations to be carried out analytically and rapidly, sidestepping the computational issues discussed in the previous section. The transition function is approximated by first propagating the mean vector and variance matrix of the CTBP from time \(t\) to \(t+1\). This calculation is relatively efficient and tractable due to the linear nature of the mean and variance operator of CTBPs. The transition function is then approximated by a multivariate Gaussian distribution with mean and variance equal to the mean and variance of the CTBP propagated to time \(t+1\). Similar approximations have a long history in stochastic modelling \citep{whittle1957} and our approach is reminiscent of moment matching \citep{lauritzen:1992} or assumed density filtering \citep[Sec 3.1]{minka:2001}.


As the filtering recursions are carried out for a fixed value of $\boldsymbol{\theta}$, we drop $\boldsymbol{\theta}$ from the expressions in this section for brevity.
\citet{dormanGardenBranchingProcesses2004} (see also Appendix \ref{APP::Branching Process Mean and Variance Calculation}) shows that mean of a CTBP over a unit time step can be written as a linear transform of the state,
\begin{equation}
    m(\textbf{z}_t)
    :=
    \E{\textbf{z}_{t + 1} \cond \textbf{z}_t} = \textbf{z}_t \textbf{F}, 
    \label{eq:mean}
\end{equation}
where the $r \times r$ matrix \(\textbf{F}\) is
\begin{equation*}
    \textbf{F} = e^{\Omega},
\end{equation*}
which can be computed using standard matrix exponential methods \citep{molerNineteenDubiousWays2003,higham2008,higham2005}. Here we use the method from \citet{higham2005} as implemented in the \texttt{exponential!}~function of the ExponentialUtilities.jl Julia package v1.27.0. 

A similar, but much longer, calculation can be performed for the variance in the state over a unit time-step (see Appendix \ref{APP::Branching Process Mean and Variance Calculation}), which yields 
\begin{equation}
     v(\textbf{z}_t)
    :=
    \text{Var}(\textbf{z}_{t + 1}\cond\textbf{z}_t)
    =
    \sum_{i = 1}^r z_{t,i} \textbf{V}_i,
    \label{eq:var}
\end{equation}
where $\textbf{V}_i$, \(i=1,...,r,\) are \(r\times r\) matrices and can be computed again using a matrix exponential. 
Together (\ref{eq:mean}) and (\ref{eq:var}) allow us to propagate the mean and variance from time \(t\) to time \(t+1\), given \(\bs z_t\). Next, we show how to propagate the mean and variance from time \(t\) to time \(t+1\), given the mean and variance of \(\bs z_t\) conditional on the observed data \(\textbf y_{1:t}\).

For the mean, by conditioning on the value of the state at time \(t\), we have 
\begin{align}
    \boldsymbol{\mu}_{t+1\cond t}&:=\E{\textbf{z}_{t + 1}\cond \textbf{y}_{1:t}} \label{eq:mt+1t}
    \\&= \E{\E{\textbf{z}_{t + 1}\cond \textbf{z}_t, \textbf{y}_{1:t}}\cond \textbf{y}_{1:t}} \nonumber
    \\&= \E{\E{\textbf{z}_{t + 1}\cond \textbf{z}_t}\cond \textbf{y}_{1:t}} \nonumber
    \\&= \E{\textbf{z}_t\textbf{F}\cond \textbf{y}_{1:t}} \tag{using (\ref{eq:mean})}
    \\&= \E{\textbf{z}_t\cond \textbf{y}_{1:t}}\textbf{F}, \nonumber
\end{align}
where the second equality uses the Markov property of the CTBP to drop the dependence on \(\textbf{y}_{1:t}\) from the inner expectation. Similarly, for the variance, using the tower property we have 
\begin{align}
    \boldsymbol{\Sigma}_{t+1\cond t}&:=\Var{\textbf{z}_{t + 1}\cond \textbf{y}_{1:t}} \label{eq:Vt+1t}
    \\&= \E{\Var{\textbf{z}_{t + 1}\cond \textbf{z}_t, \textbf{y}_{1:t}}\cond \textbf{y}_{1:t}} + \Var{\E{\textbf{z}_{t + 1}\cond \textbf{z}_t, \textbf{y}_{1:t}}\cond \textbf{y}_{1:t}} \nonumber
    \\&= \E{\sum_{i=1}^r\textbf{V}_i z_{i,t}\cond \textbf{y}_{1:t}} + \Var{\textbf{z}_{t} \mathbf F \cond \textbf{y}_{1:t}} \tag{using (\ref{eq:var})}
    \\&= \sum_{i=1}^r\textbf{V}_i \E{z_{i,t}\cond \textbf{y}_{1:t}} + \textbf{F}\Tr \Var{\textbf{z}_{t}\cond \textbf{y}_{1:t}} \textbf{F}, \nonumber
\end{align}
where, again, we have used the Markov property to drop the dependence on \(\mathbf y_{1:t}\) in the inner variance and expectation operators from the second to the third line. Notice that the mean and variance of the predictive distributions are simply matrix transforms of the mean and variance of \(\textbf{z}_t\cond \textbf{y}_{1:t}\). Appendix \ref{APP::Branching Process Mean and Variance Calculation} details how the matrices \textbf{F} and \textbf{V}$_i$ may be calculated efficiently as matrix exponentials. 

At this point we perform the approximation and approximate the distribution of \(\textbf{z}_{t + 1}\cond \textbf{y}_{1:t}\) as a Gaussian distribution using the mean \(\bs \mu_{t+1\cond t}\) and variance \(\boldsymbol{\Sigma}_{t+1\cond t}\). Since the observation distribution is a Gaussian distribution, then the usual Kalman filter update step can be used to calculate the posterior distribution of the state at time \(t+1\), \(p(\textbf{z}_{t+1}\cond \textbf{y}_{1:t+1})\) from which the mean and variance can be easily obtained, and this completes one step of the filtering recursion. 


Approximating the transition function with a Gaussian distribution is convenient as it enables the integrals in the filter to be carried out analytically and rapidly as matrix operations. There are some issues with this choice, however. For one, the approximation only captures the first two moments of the transition function. If the higher-order moments differ significantly from those of the Gaussian approximation then the approximation may not give accurate results. Further, the Gaussian approximation is supported on \(\mathbb R^r\), but the transition function takes values in \(\mathbb N^r\) only. Hence, this difference may cause the approximation to give inaccurate results when the approximation puts significant probability on the set \(\{(z_1,z_2,...,z_r)\cond z_i<0 \text{ for some }i=1,...,r\}\).

The procedure for a single step of the approximate filtering algorithm is given in Appendix~\ref{SEC::Filtering Distribution}. Implementation of the approximate filtering algorithm has two main components, calculation of the matrices \(\textbf{F}\) and \(\textbf{V}_i,\,i=1,...,r\), and applying the Kalman filter steps (this is equivalent to a usual Kalman filter with a time-inhomogeneous movement model variance). The matrices \(\textbf{F}\) and \(\textbf{V}_i,\,i=1,...,r\), need to be computed once for a given set of CTBP parameters. For example, in Section~\ref{SEC::Victoria Second COVID-19 Wave} we apply the approximate algorithm to a CTBP model with time-inhomogeneous parameters which are piecewise constant and change every 7 days; there are 98 days of data in total and hence fourteen sets of parameters, so we need to compute the matrices \(\textbf{F}\) and \(\textbf{V}_i,\,i=1,...,r\), fourteen times to evaluate the likelihood once. As shown in Appendix~\ref{APP::Branching Process Mean and Variance Calculation}, the calculation of these matrices amounts to computing the matrix exponential of an \(r(r+1)\times r(r+1)\) matrix, which is computationally feasible for a moderate number of types \(r\). Recall, within a Metropolis Hastings scheme the likelihood typically needs to be evaluated many times (of the order \(1e5-1e6\)) so we need the evaluation of the likelihood to be rapid. For the model in Section~\ref{SEC::Victoria Second COVID-19 Wave}, the computation of \(\textbf{F}\) and \(\textbf{V}_i,\,i=1,...,r\), takes the majority of the total computation time.

\subsection{Switching between Filtering Methods}
\label{SEC::Switching from PF to KF}
In this section, we discuss how the filtering procedure can switch between the Gaussian approximation or a particle filter (PF) based on the estimated agent count at each observation time. For the application in mind, the typical situation is one in which the method uses a PF initially and later switches to the approximate Gaussian filter once the population size become large enough.


Define $\mathcal{T}$ as the set of times where the approximation will be used. Suppose $\tau \not\in \mathcal{T}$ but $\tau+1 \in \mathcal{T}$, so there is a switch from the PF to the Gaussian approximation between \(\tau\) and \(\tau+1\). To apply the Gaussian approximation we need the mean and variance of the state posterior distribution, \(p(\textbf{z}_{\tau}\cond \textbf{y}_{1:\tau})\). The distribution of \(\textbf{z}_{\tau}\cond \textbf{y}_{1:\tau}\) is approximated by a set of \(n\) samples from the PF (particles), $\{\textbf{z}^{(1)}_\tau, \dots , \textbf{z}^{(n)}_\tau\}$. We estimate the mean and variance of \(\textbf{z}_{\tau}\cond \textbf{y}_{1:\tau}\) using the particles by 
\begin{equation*}
    \label{EQN::Mean & Covariance calc}
    \hat{\boldsymbol{\mu}}_{\tau\cond\tau} := \frac{1}{n}\sum_{k = 1}^n \textbf{z}^{(k)}_\tau,
    \quad
    \hat{\boldsymbol{\Sigma}}_{\tau\cond\tau} := \frac{1}{n}\sum_{k = 1}^n (\textbf{z}_\tau^{(k)} - \hat{\boldsymbol{\mu}}_{\tau\cond\tau})^T(\textbf{z}_\tau^{(k)} - \hat{\boldsymbol{\mu}}_{\tau\cond\tau}).
\end{equation*}
With these estimates, the Gaussian approximation can then be applied.

Now suppose that $\tau \in \mathcal{T}$ but $\tau+1 \not\in \mathcal{T}$, so there is a switch from the Gaussian approximation to the PF between \(\tau\) and \(\tau+1\). In this case, we can simply draw $n$ rounded and censored (so that the samples have non-negative entries) samples from $\mathcal{N}(\boldsymbol{\mu}_{\tau\cond\tau}, \boldsymbol{\Sigma}_{\tau\cond\tau})$, assign them equal weights, and proceed filtering with the PF. 

In either case, the overall estimate of the marginal likelihood is 
\begin{eqnarray}
    \hat{p}(\textbf{y}_{1:T}\cond\boldsymbol{\theta}) =
    \prod_{t\in \mathcal{T}} \phi(\textbf{y}_{t};\textbf{H}\boldsymbol{\mu}_{t| t - 1}\Tr, \textbf{H}\boldsymbol{\Sigma}_{t| t - 1}\textbf{H}\Tr+\textbf{R} ) \prod_{t\in \set{1,2,\dots , T}\setminus\mathcal{T}}
    \hat{p}(\mathbf{y}_{t}\cond\mathbf{y}_{1:t - 1}),
    \label{EQN::Y_{1:T}| theta (approx)} 
    \label{eq:switching_approx}
\end{eqnarray}
where $\hat{p}(\mathbf{y}_{t}\cond\mathbf{y}_{1:t - 1})$ is the likelihood estimate obtained from the $t^\text{th}$ step of the PF, \(\phi(\textbf{y}_{t};\boldsymbol{\mu}, \boldsymbol{\Sigma})\) is the density of a multivariate Gaussian distribution with mean \(\boldsymbol{\mu}\) and covariance \(\boldsymbol{\Sigma}\) evaluated at the observation \(\textbf{y}_{t}\), \(\boldsymbol{\mu}_{t| t - 1}\) and \(\boldsymbol{\Sigma}_{t| t - 1}\) are defined in (\ref{eq:mt+1t}) and (\ref{eq:Vt+1t}), respectively, and \(\bs H\) and \(\bs R\) are the parameters of the observation model.

Defining the set $\mathcal{T}$ is a trade-off between computational time and accuracy.  Unlike the PF, the time complexity of a single step of the Gaussian filter is independent of the number of samples being used and the total agent population. Thus, the overall time complexity of a single run of the algorithm to calculate Equation~\eqref{eq:switching_approx} is well approximated by the complexity of the steps when the PF is used. This will be $\mathcal{O}\br{n \sum_{t\in \mathcal{T}}\mathcal{E}_{t}}$, where $n$ is the total number of particles (assumed constant over all time steps) and $\mathcal{E}_t$ is the expected number of simulated events that occur over the interval $[t - 1, t)$. 

The efficiency of the PMMH chain is sensitive to the variance in the likelihood estimate and so this needs to be controlled \citep{pittPropertiesMarkovChain2012,doucetEfficientImplementationMarkov2015,sherlockEfficiencyPseudomarginalRandom2015}.
The expected number of simulated events is proportional to the number of agents at time $t$, which will, in our epidemic settings, typically grow exponentially with time and is not a controllable parameter. In practice the value of $n$ must be tuned to obtain sufficiently low variance. If the Gaussian approximation is used more often ($|\mathcal{T}|$ is larger), the variance will naturally be lower as no simulation is taking place during these steps, but this may decrease the accuracy of the posteriors if the approximation is not valid. This leads to a trade-off between the accuracy of the posteriors (for both the states and parameters) and computational expense, which we control by setting a threshold to determine the size of the set $\mathcal{T}$. 

In this paper, we choose $\mathcal{T}$ based on the mean agent counts at time $t$;
\begin{equation}
    \mathcal{T} :=  \set{
        t \in 1,2,\dots ,T : \min(\hat{\boldsymbol{\mu}}_{t \cond t}) \geq s}, \label{EQN::Switching threshold min comp}
\end{equation}
where $\min(\hat{\boldsymbol{\mu}}_{t \cond t})$ is the minimum over all the elements of the mean vector and $s$ is a pre-chosen threshold parameter. Thus, the set \(\mathcal T\) is determined dynamically as the algorithm is run. This formulation has the advantage that it is closely related to the assumption that the state distribution is approximately a Gaussian distribution when the agent counts are large. However, making $\mathcal{T}$ dependent on state estimates may cause issues if the Gaussian approximation is significantly biased in the calculation of the likelihood, as this could cause certain state estimates to be artificially favoured. Another possibility, which we do not investigate here, is to set $\mathcal{T}$ based solely on the observations, $\mathbf{y}_{t},\, t=1,...,T$ which has the advantage that $\mathcal{T}$ can be pre-determined but does not take into account how the dynamics change with the model parameters, $\boldsymbol{\theta}$. For example, for some values of $\boldsymbol{\theta}$ the populations will grow slower, in which case it would be better to use the PF for longer before switching to the approximation. 



Smoothed state posterior distributions, \(p(\mathbf z_t\cond \mathbf y_{1:T})\), can be obtained from the output of the Gaussian approximation using the Rauch-Tung-Striebel smoother \citep{sarkkaBayesianFilteringSmoothing2013}. To compute the smoothed states posterior distribution for $t\not\in \mathcal{T}$, the algorithm must track the resampling indexes and particle weights so trajectories can be reconstructed backwards through time \citep{chopinIntroductionSequentialMonte2020}.

\section{Results}
\label{SEC::Results}
In this section we compare the efficiency, state estimates and parameter estimates produced using the PF and our Gaussian and hybrid approximations. We use the PF and the approximation methods to evaluate the likelihood within a Metropolis-Hastings scheme to sample from the posterior distributions. Using the PF, the samples are from the true posterior distribution (assuming the chain is sufficiently mixed). Hence, we can validate the accuracy of the approximations by comparing the posterior distributions they produce with those produced by the PF. The purpose of the approximations is to produce an alternative to the PF which is significantly faster but with minimal loss in accuracy. In Sections~\ref{SEC::SEIR Branching Process} and \ref{SEC::SE8I8R Branching Process} we test SEIR (Susceptible-Infectious-Recovered) and SE8I8R epidemic models  respectively \citep{black:2009}. Further numerical experiments for the same models are documented in Appendix~\ref{APP: further sims}. Section \ref{SEC::Victoria Second COVID-19 Wave} applies the approximation to case data from the Second COVID-19 Wave in Victoria, Australia.

\subsection{SEIR Branching Process}
\label{SEC::SEIR Branching Process}
Here we investigate the characteristics of the proposed likelihood evaluation methods, the Gaussian approximation, particle filter, and the hybrid algorithm with a switching threshold of \(s=10\), for the simple SEIR example introduced earlier.

\begin{figure}
    \centering
    \includegraphics[width=0.9\textwidth]{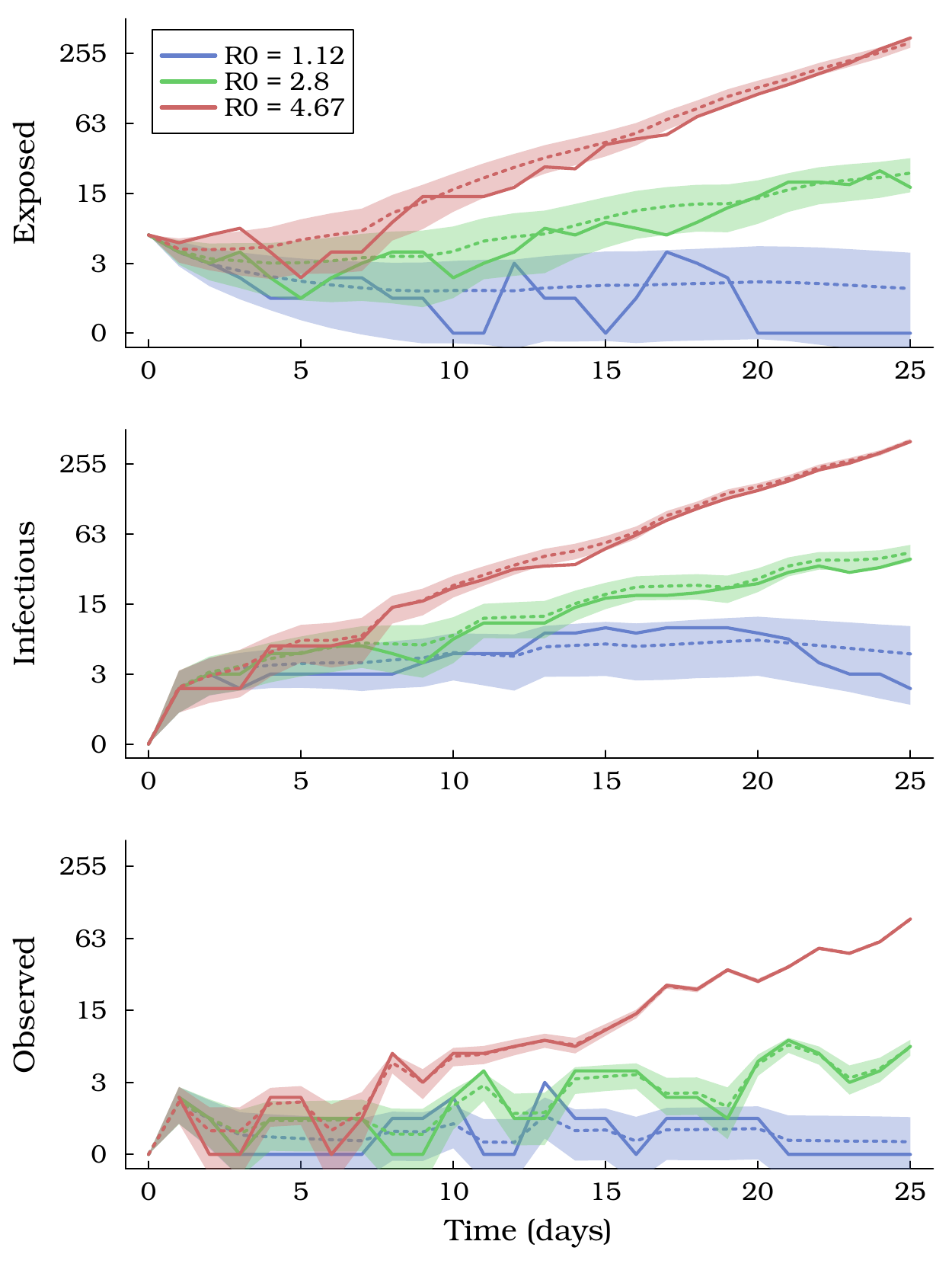}
    \caption{Simulated realisations of the SEIR epidemic model for three values of the reproductive number \(R_0=\beta/\lambda=1.12,\,2.8,\,4.\overline{6}\) (solid lines) and the median (dashed lines) and symmetric 80\% credible interval of the filtering distribution, \(p(z_{t,i}\cond \mathbf y_{1:t}),\, i=1,2,3\), estimated by the Gaussian approximation. Note, the y-axes have logarithmic scales and that the filtering distributions are only estimated from the observed case data, so the exposed and infectious are unobserved.}
    \label{FIG:Simulated SEIR data}
\end{figure}

To generate data to fit our models to, we fix \(p=0.75,\, \delta=0.375,\, \lambda=3/28\approx 0.1071,\,\sigma^2=0\) and initial state \(\bs z_0=(6,0,0),\) then generate realisations of \(\bs z_t\) for three values of the reproductive number \(R_0=\beta/\lambda=1.12,\, 2.8,\, 4.\overline{6}\) (\(\beta=0.12,\,0.3,\,0.5\)) corresponding to slow, medium and fast spreading epidemics.
Note that the synthetic data for this model, and the next, is generated assuming that there is no additional noise, i.e.~$\mathbf{R}$ is a matrix of zeros. However, when performing inference, we assume non-zero variance as this improves the behaviour of the particle filters that we use for exact sampling results. The effect of this regularisation is explored in results shown in Appendix~\ref{APP: further sims}.
The simulated data is shown in Figure~\ref{FIG:Simulated SEIR data}. We fit our model to the data shown in the bottom plot in Figure~\ref{FIG:Simulated SEIR data} only; the number of exposed and infectious individuals is unobserved.


We use our methods to infer the posterior distribution of the reproductive number, \(R_0\), in a Metropolis-Hastings scheme, fixing all other parameters at their true values. This is because with only case data available, only a single parameter of the model will be identifiable. Multiple parameters can be inferred but the resulting marginal posteriors will be wider and this distracts from our aim of investigating the performance of the approximation.  Throughout this section, the prior distribution on the reproductive number is \(R_0\sim \mbox{Gamma}(4.4,0.5)\) (which has mean \(2.2\)). In all cases the Metropolis-Hastings chain is run for 81920 steps; the first 20480 samples are used to adapt the proposal distribution and are then discarded as burn-in. The adaptive scheme sets the variance of the proposal distribution of the chain (a Normal distribution centred at the current sample) to a scalar multiple of the variance computed from the previous 4096 samples of the chain. Except for changing how the likelihood is evaluated, the same Metropolis-Hastings scheme is used in all cases. The particle filter and hybrid methods both use 256 particles and all computations are performed with 1 thread.

\subsubsection*{Changing the reproductive number}
\begin{figure}
\begin{minipage}{0.5\textwidth}
    \centering
    Vary \(R_0\)
    \includegraphics[width=\textwidth]{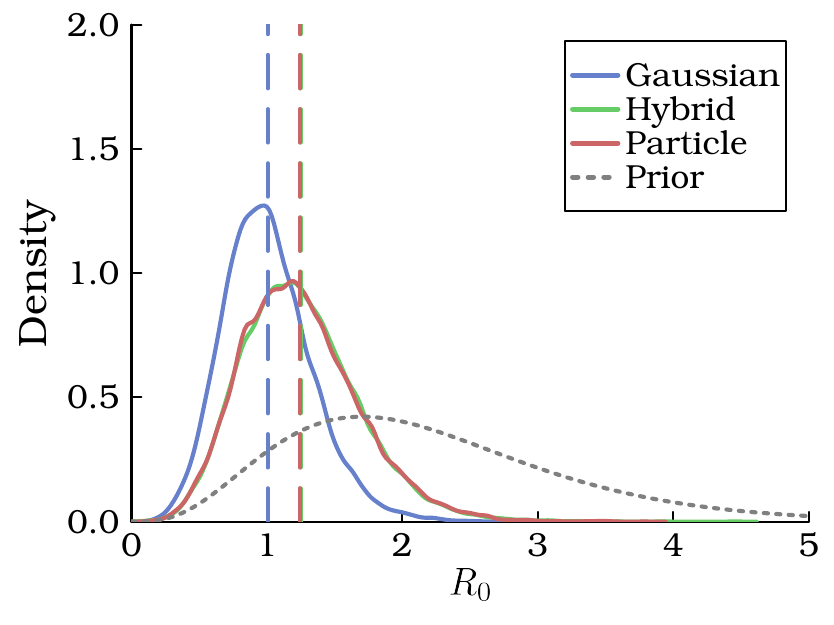}
    \includegraphics[width=\textwidth]{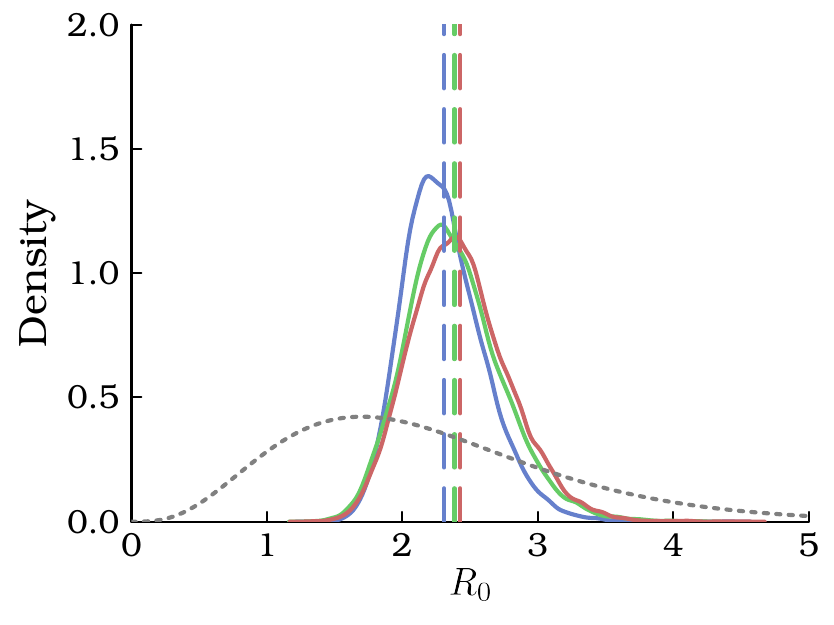}
    \includegraphics[width=\textwidth]{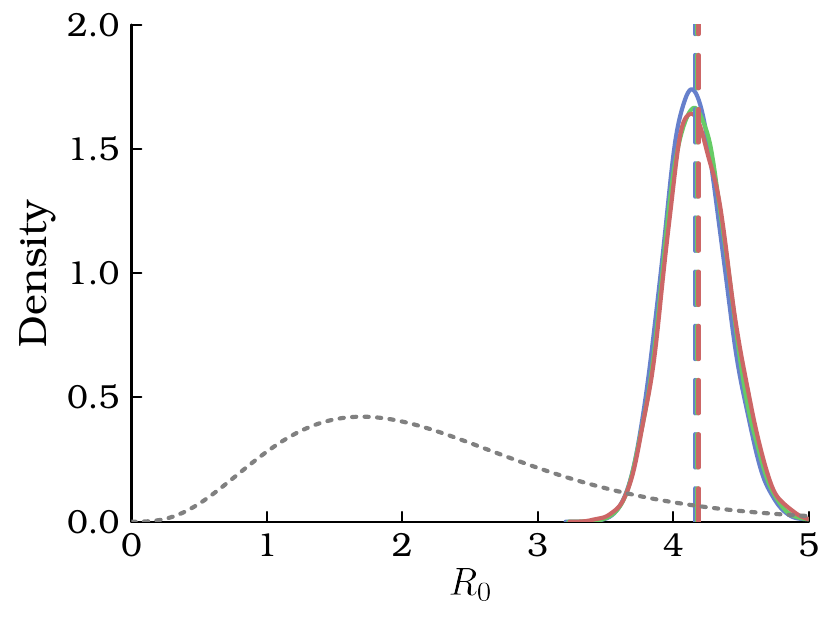}
\end{minipage}%
\begin{minipage}{0.5\textwidth}
    \centering
    Vary \(T\)
    \includegraphics[width=\textwidth]{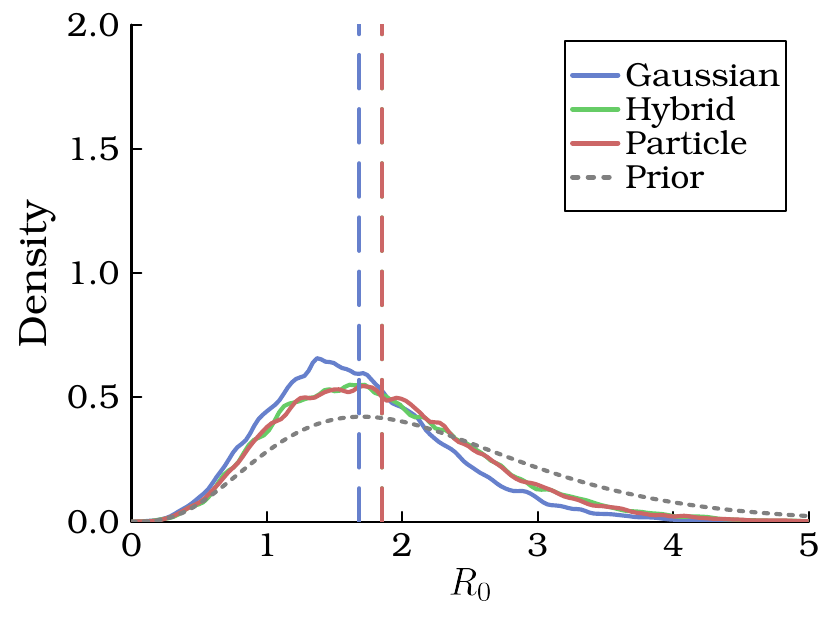}
    \includegraphics[width=\textwidth]{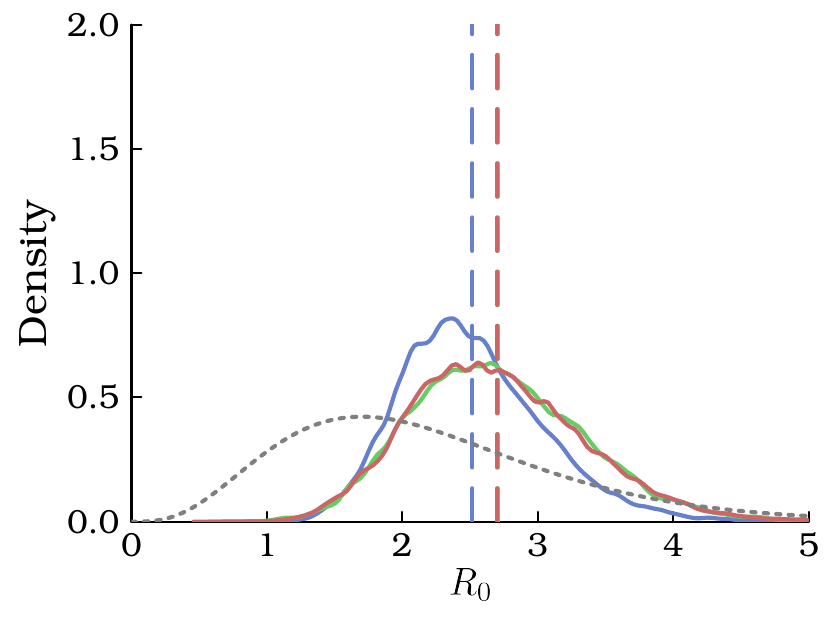}
    \includegraphics[width=\textwidth]{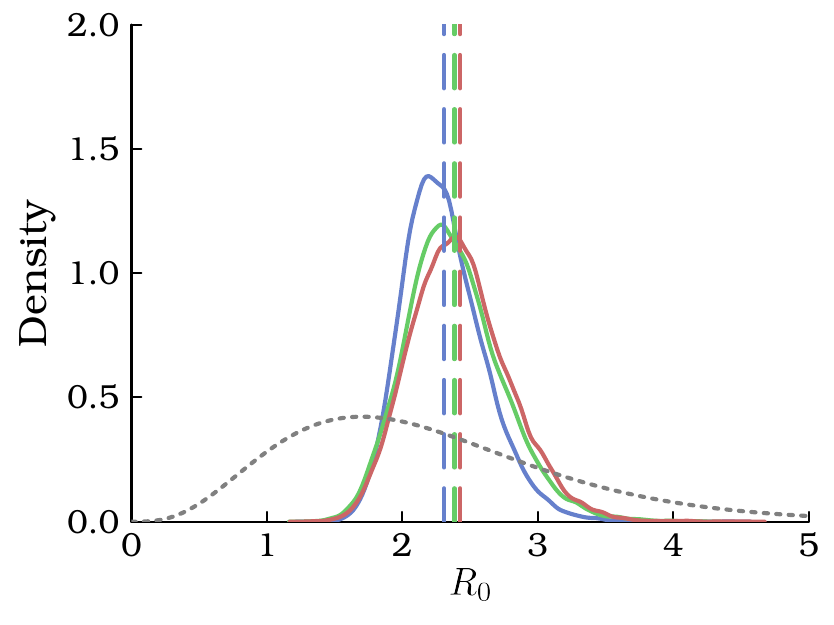}
\end{minipage}
\caption{\label{FIG:Vary R0 and T SEIR}(Left) Posterior distributions for three different values of the reproductive number \(R_0=\beta/\lambda=1.12,\,2.8,\,4.\overline{6}\), (top, middle and bottom, respectively) and \(T=25\) using the three methods to evaluate the likelihood for the SEIR model. (Right) Posterior distributions for three different lengths of time-series \(T=10,\,15,\,25\), (top, middle and bottom, respectively) and \(R_0=2.8\) using the three methods to evaluate the likelihood for the SEIR model. The vertical lines are at the mean of the posterior distributions.}
\end{figure}

In this section we fix \(T=25\) and perform inference on the three difference time series (generated with different $R_0$ values) while holding all other parameters fixed at their true values. Figure~\ref{FIG:Vary R0 and T SEIR}~(left) shows the posterior distributions for using the three methods for the three different values of \(R_0\). These demonstrate that the Gaussian approximation is biased towards lower values of \(R_0\) for \(R_0=1.12\), is comparatively less biased for \(R_0=2.8\), and shows very little bias for \(R_0=4.\overline{6}\), where the populations grow more quickly. 

The Hybrid method has relatively little bias for \(R_0=1.12\), which is likely because the switching threshold is rarely reached and the particle filter is used for the majority of time-steps. 
The Hybrid method displays relatively more bias for \(R_0=2.8\) compared to the \(R_0=1.12\) case. This is likely caused by the switching threshold being exceeded more often, for \(R_0=2.8\) than for \(R_0=1.12\). For \(R_0=2.8\) the posterior distribution is less biased than the Gaussian approximation, suggesting that the switching is beneficial. 

Table~\ref{TAB::Vary R0 SEIR} shows the effective samples size (ESS) and run times for the Metropolis-Hastings sampler for the three different methods and three values of \(R_0\). Run time, ESS and ESS/sec are calculated excluding burn in. The ESS is a proxy for how well the Metropolis-Hastings chain samples the posterior, with a higher ESS being better. As shown in Table~\ref{TAB::Vary R0 SEIR}, the ESS is similar in all cases except the particle filter in the case \(R_0=4.\overline{6}\). We suggest that this is due to a large variance in the likelihood estimate calculated using the particle filter which cases degradation in the sampler. ESS/sec is significantly greater for the Gaussian approximation compared to the PF.

\begin{table}
    \centering 
    \begin{tabular}{ccrrr}$R_0$ & Method & ESS & ESS/sec & Time (s)\\ \hline \hline 
1.12 & Gaussian & 12864.5 & 10075.7 & 1.3 \\ 
1.12 & Hybrid & 11358.0 & 68.5 & 165.8 \\ 
1.12 & Particle & 11346.7 & 68.4 & 165.8 \\ \hline 
2.8 & Gaussian & 12726.6 & 6792.1 & 1.9 \\ 
2.8 & Hybrid & 9631.1 & 42.6 & 225.8 \\ 
2.8 & Particle & 9161.7 & 16.1 & 569.9 \\ \hline 
$4.\overline 6$ & Gaussian & 12984.9 & 7539.8 & 1.7 \\ 
$4.\overline 6$ & Hybrid & 11297.9 & 87.7 & 128.9 \\ 
$4.\overline 6$ & Particle & 4690.0 & 1.6 & 2956.9 \\ \hline 

    \end{tabular}
    \caption{Table of effective sample sizes and run times for the three different values of the \(R_0=1.12,\, 2.8,\, 4.\overline{6}\), using the three methods to evaluate the likelihood for the SEIR model.}
    \label{TAB::Vary R0 SEIR}
\end{table}

\begin{figure}
\begin{minipage}{0.5\textwidth}
    \centering
    \(R_0=1.12\)
    \includegraphics[width=\textwidth]{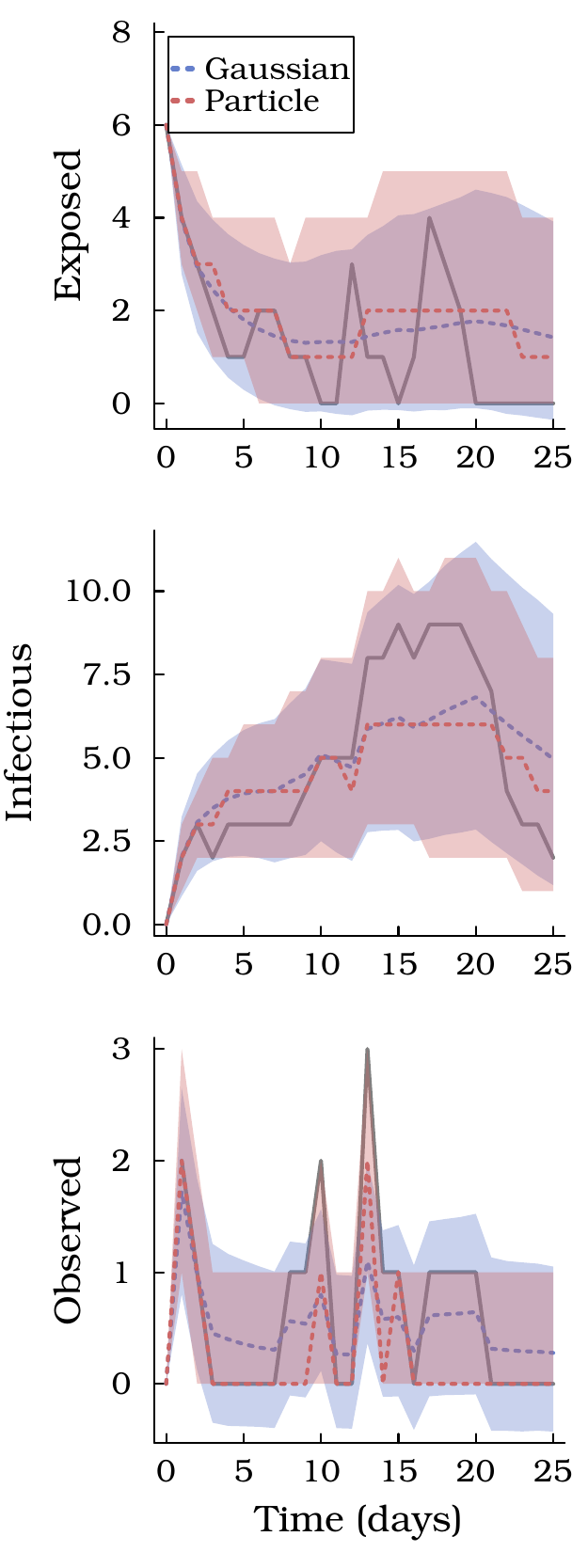}
\end{minipage}%
\begin{minipage}{0.5\textwidth}
    \centering
    \(R_0=4.\overline{6}\)
    \includegraphics[width=\textwidth]{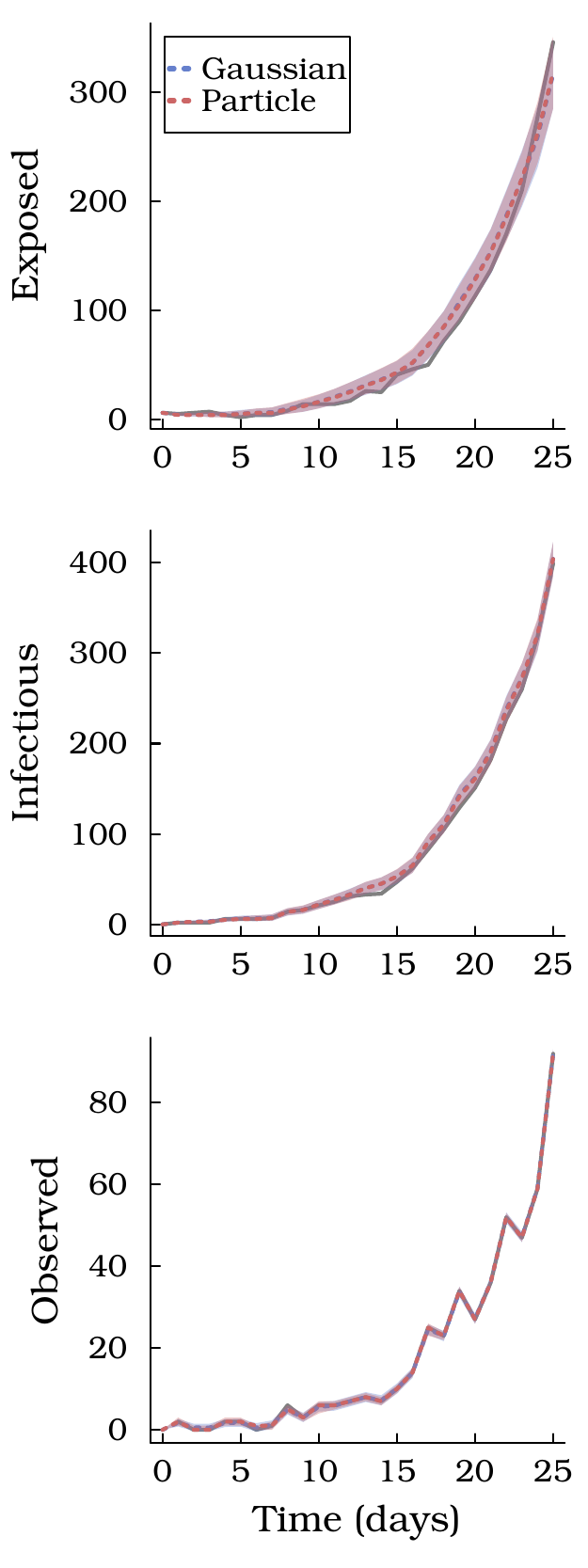}
\end{minipage}
\caption{\label{FIG:filtered SEIR data}The median and symmetric 80\% credible interval of the filtering distribution, \(p(z_{t,i}\cond \mathbf y_{1:t}),\, i=1,2,3\), estimated by the Gaussian approximation (blue) and PF (red), and the simulated realisations of the SEIR epidemic model for \(R_0=\beta/\lambda=1.12\) (left) and \(4.\overline{6},\) (right) (solid grey lines).}
\end{figure}

Figure~\ref{FIG:filtered SEIR data} shows the median and symmetric 80\% credible interval of the filtering distribution, \(p(z_{t,i}\cond \mathbf y_{1:t}),\, i=1,2,3\), estimated by the Gaussian approximation and PF for the cases \(R_0=1.12\) and \(R_0=4.\overline{6}\) (the case with \(R_0=2.8\) is in Figure~\ref{FIG:filtered p/var SEIR data}) and demonstrates good agreement between the quantiles of the approximation and true filtering distributions.

Figure~\ref{FIG:Simulated SEIR data} shows the median and symmetric 80\% credible interval of the filtering distribution, \(p(z_{t,i}\cond \mathbf y_{1:t}),\, i=1,2,3\), estimated by the Gaussian approximation for \(R_0=1.12,\, 2.8\) and \(4.\overline{6}\). This plot suggests that the Gaussian approximation is reasonable since the simulated data typically lies in the credible intervals.

\subsubsection*{Changing the amount of data}
Now, fix \(R_0=2.8\) and let \(T=10,\,15,\,25\) vary, while holding all other parameters fixed at their true values. Figure~\ref{FIG:Vary R0 and T SEIR}~(right) shows the posterior distributions for using the three methods for the three different values of \(T\).

Figure~\ref{FIG:Vary R0 and T SEIR}~(right) demonstrates that the Gaussian approximation is consistently biased towards lower values of \(R_0\) in all three cases. There is no strong evidence that the bias is better, or worse, as \(T\) increases. The posterior distributions produced by the hybrid method are close to the true posterior produced using the particle filter.

Table~\ref{TAB::Vary T SEIR} shows evidence that the ESS for the PF degrades as the amount of data, \(T\), increases. In comparison, the ESS for the Gaussian approximation does not appear to decrease with \(T\). The run time of the Gaussian approximation is much faster than the other methods for this model.

\begin{table}
    \centering 
    \begin{tabular}{ccrrr}$T$ & Method & ESS & ESS/sec & Time (s)\\ \hline \hline 
10 & Gaussian & 13086.2 & 11038.7 & 1.2 \\ 
10 & Hybrid & 12350.4 & 173.2 & 71.3 \\ 
10 & Particle & 11216.5 & 157.7 & 71.1 \\ \hline 
15 & Gaussian & 12534.7 & 10196.1 & 1.2 \\ 
15 & Hybrid & 11380.6 & 68.2 & 166.9 \\ 
15 & Particle & 10786.4 & 51.9 & 208.0 \\ \hline 
25 & Gaussian & 12726.6 & 6792.1 & 1.9 \\ 
25 & Hybrid & 9631.1 & 42.6 & 225.8 \\ 
25 & Particle & 9161.7 & 16.1 & 569.9 \\ \hline 

    \end{tabular}
    \caption{Table of effective sample sizes and run times for the three different values of the \(T=10,15,25\), using the three methods to evaluate the likelihood for the SEIR model. Run time, ESS and ESS/sec are calculated excluding burn in.}
    \label{TAB::Vary T SEIR}
\end{table}

\subsubsection*{Posterior distribution of the state of the branching process}
\begin{figure}
    \begin{minipage}{0.5\textwidth}
    \centering
    Exposed (State 1)
    \includegraphics[width=\textwidth]{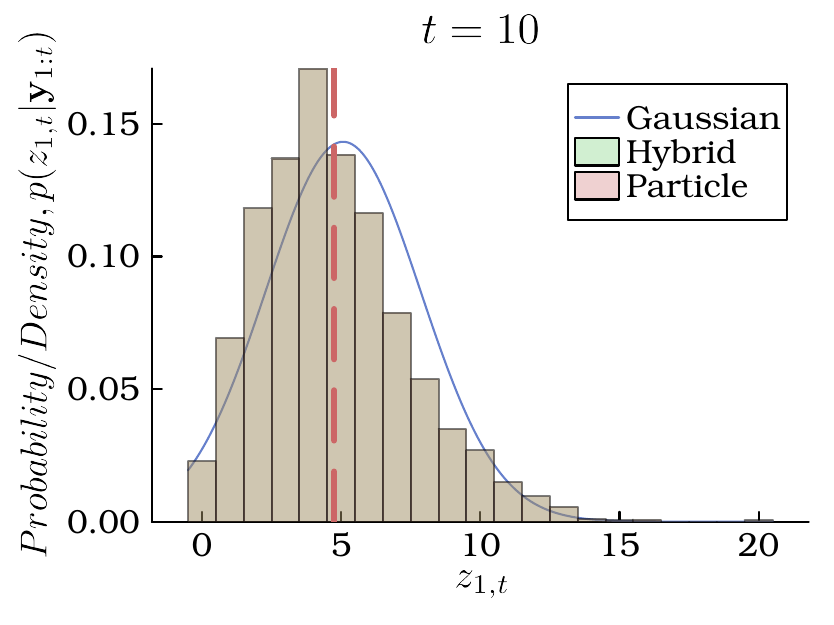}
    \includegraphics[width=\textwidth]{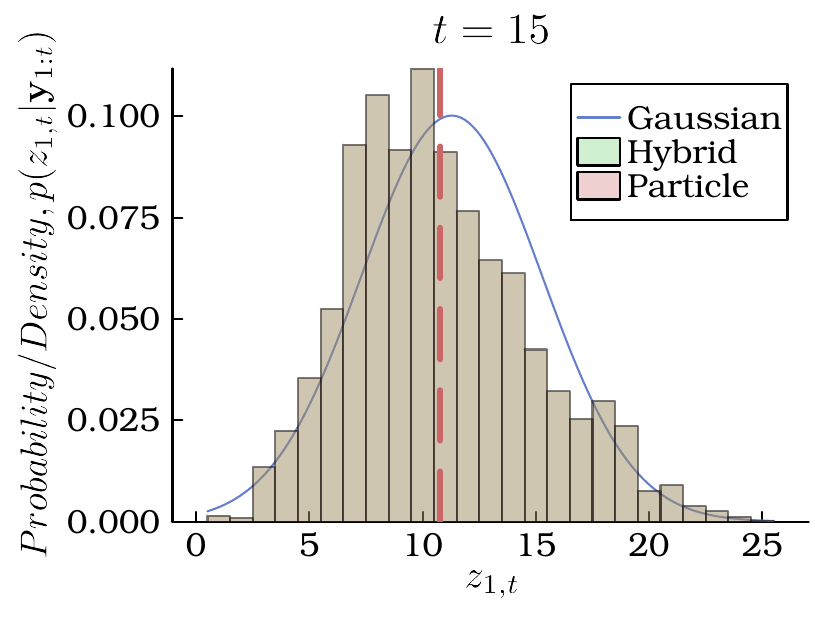}
    \includegraphics[width=\textwidth]{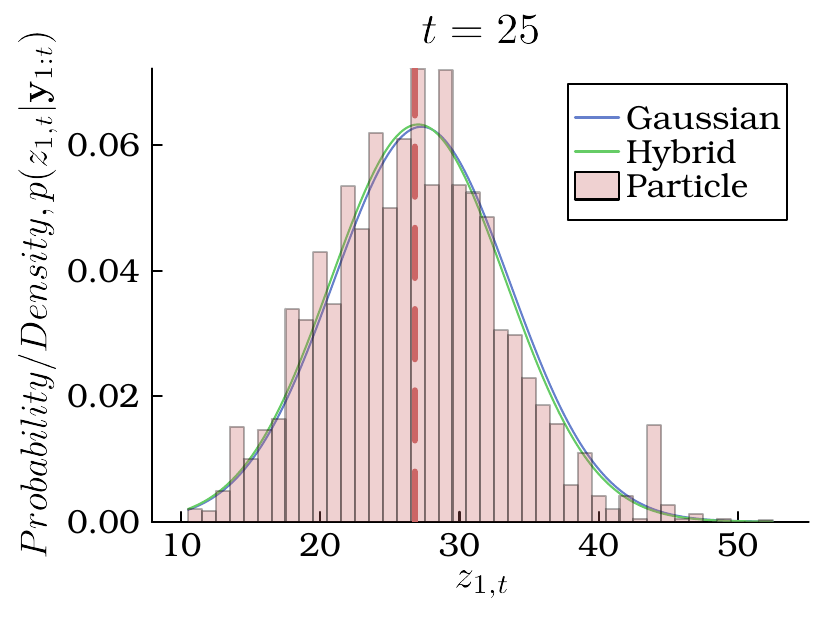}
    \end{minipage}%
    \begin{minipage}{0.5\textwidth}
    \centering
    Infectious (State 2)
    \includegraphics[width=\textwidth]{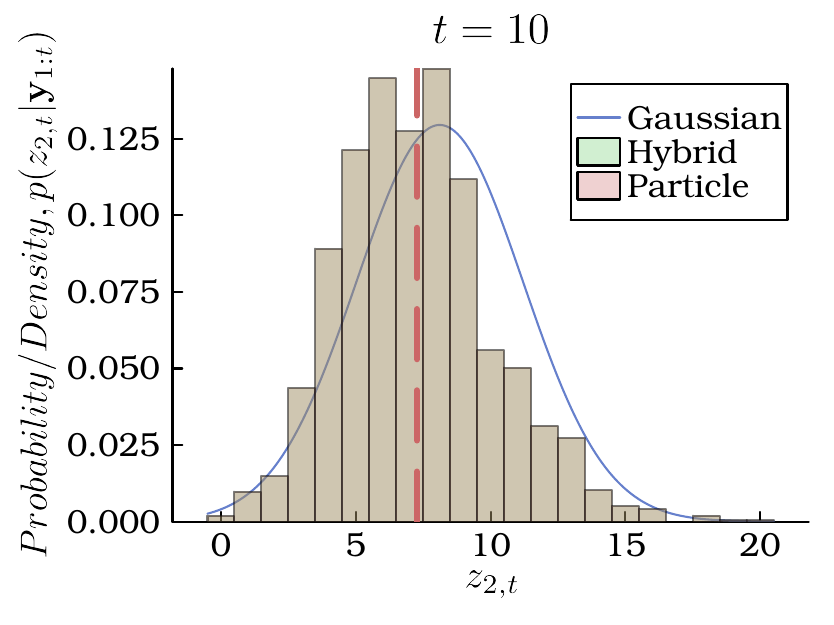}
    \includegraphics[width=\textwidth]{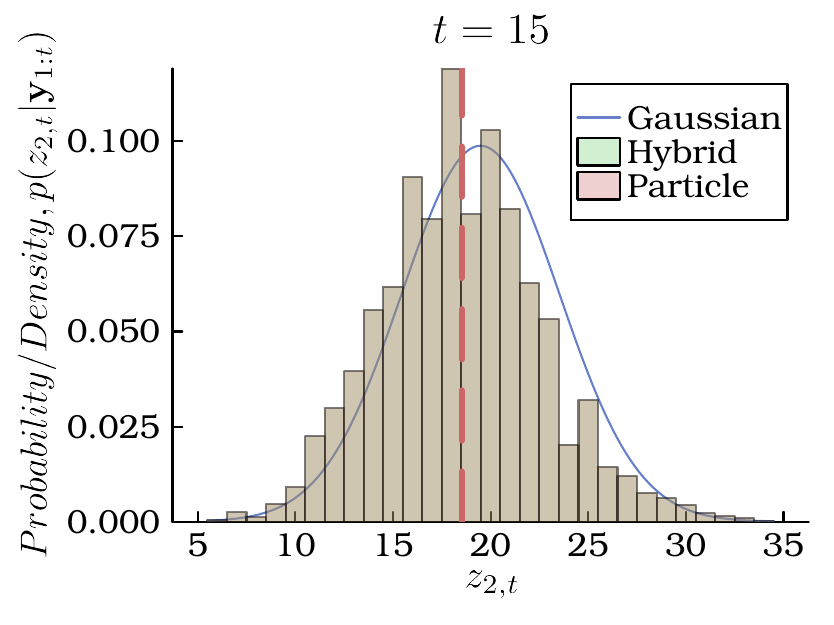}
    \includegraphics[width=\textwidth]{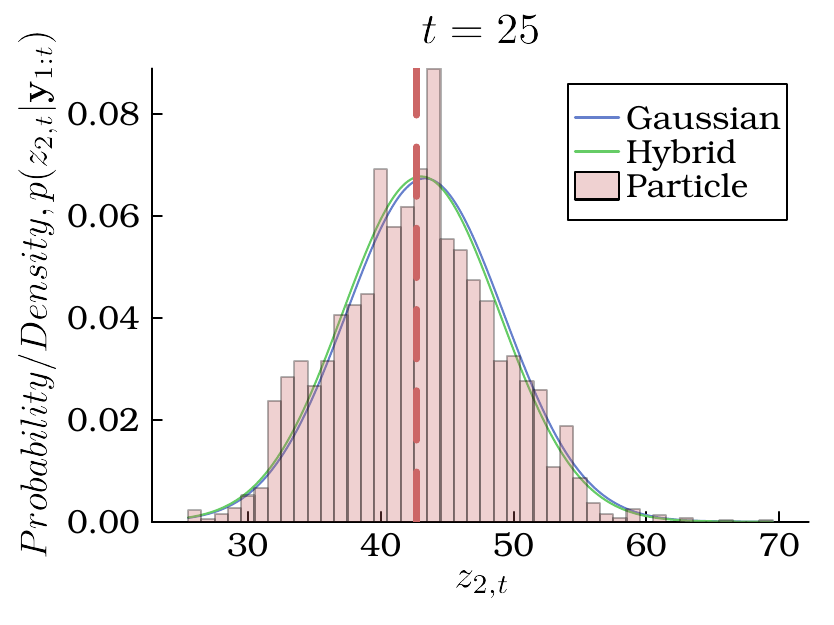}
    \end{minipage}
    \caption{\label{FIG:state posteriors SEIR} Posterior distributions of the counts of Exposed (left) and Infectious (right) individuals at times \(T=10,15,25\) (top, middle, bottom, respectively) for the SEIR model. Distributions approximated by particles are represented as histograms, else they are represented as densities. The vertical line is at the mean of the particles from the particle filter and estimates the true mean.}
\end{figure}

The posterior distributions of the state of the branching process \(p(\bs z_{t}\cond \bs y_{1:t})\) with the parameters fixed at their true values and \(R_0=2.8\) are shown in Figure~\ref{FIG:state posteriors SEIR}. At times \(t=10\) and \(t=15\) the posterior distributions produced by the hybrid and particle methods are the same because the switching threshold has not-yet been crossed, and at \(t=25\) the posterior distributions produced via the hybrid and particle methods differ. Figure~\ref{FIG:state posteriors SEIR} demonstrates that the posterior distributions produced by the approximations (hybrid and Gaussian) are reasonable, and appear to get better at larger times \(t\) with larger population sizes.

\subsubsection*{Other experiments}

Appendix \ref{APP: further sims} presents results of some further numerical experiments to investigate the effects of changing the observation probability and variance for both models. 

\subsection{SE8I8R Branching Process}
\label{SEC::SE8I8R Branching Process}
Now we investigate the characteristics of the proposed likelihood evaluation methods for an SE8I8R model. This model is the same as the SEIR model, except individuals who become infected transition from susceptible to infectious where they progress through eight exponentially distributed Exposed states, \(E_1\to E_2 \to ... \to E_8\) (all with the same rate); then from exposed to infectious where they then progress through eight exponentially distributed Infectious states, \(I_1\to I_2 \to ... \to I_8\) (all with the same rate). This splitting of the periods into stages gives the overall distribution of time spent in them an Erlang (or Gamma) distribution, thus potentially representing the biology of the process better \citep{lloyd:2001a,black:2009}.  For our purpose here, this example is to compare the particle filter and the approximation when the dimension of the unobserved latent states is higher. With a larger number of types, on average, we expect the populations of each type to be much lower, hence this case tests the approximation in a setting with low case number of each type. The dimension of the characteristic matrix and matrices in the Kalman filter are also larger, hence this case also investigates possible complications which may arise from these larger matrices. 

The events and parameters of the SE8I8R branching process model are summarised in Table~\ref{TAB::EX2 Events and state transition}. Additionally, there is also the parameter for the variance of the observations, \(\sigma^2\), and the length of the time-series (number of observations), denoted \(T\). Here, we fit the SE8I8R model to the same data we did before in Section~\ref{SEC::SEIR Branching Process}. To fit the model, unless otherwise stated, we set \(p=0.75,\,\sigma^2=1,\) and \(\delta=0.375\times 8 = 3.0,\, \lambda=3/28\times 8\approx 0.8571,\) so that the overall mean exposed and infectious periods are the same as before; the initial state is \(\bs z_0=(1,1,1,1,1,1,0,0,0,0,0,0,0,0,0,0,0),\) (a total of 6 exposed and no infectious individuals).

\begin{table}
    \centering 
    \begin{tabular}{lc}
        \hline
        Event & Rate \\
        \hline
        \hline 
        \(S\to E_1\) & $\beta \sum_{i=1}^8I_{i,t} $  \\
        \(E_{i}\to E_{i+1},\,i=1,...,7\)& $\delta E_{i,t} $  \\
        \(E_8\to I_1\) (observed)& $p\delta E_{8,t} $  \\
        \(E_8\to I_1\) (unobserved)& $(1-p)\delta E_{8,t} $  \\
        \(I_{i}\to I_{i+1},\,i=1,...,7\)& $\lambda I_{i,t} $  \\
        Removal & $\lambda I_{8,t}$  \\
        \hline
    \end{tabular}
    \caption{Summary of events and rates for the SE8I8R model as a birth-death process.}
    \label{TAB::EX2 Events and state transition}
\end{table}

For the SE8I8R model, since there are 8 times as many exposed and infectious stages to progress through than the SEIR model, and since the mean exposed and infectious periods are the same, then 8 times as many events occur as an individual progresses from susceptible to recovered, increasing the computational burden in the particle filter. The computational burden in the Gaussian approximation is also greatly increased by the increase in the number of types in the model. In Appendix~\ref{APP::Branching Process Mean and Variance Calculation} Equation~(\ref{EQN:: var matrix exp}) we show that the variance of the branching process can be calculated by computing a matrix exponential of dimension \(r(r+1) \times r(r+1)\) where \(r\) is the number of types. For the SEIR model there are 3 types, so this matrix is \(12\times 12\) (144 entries); for the SE8I8R model there are 17 types, so this matrix is \(306\times 306\) (93,636 entries). 


The same Metropolis-Hastings schemes as in Section~\ref{SEC::SEIR Branching Process} was used to sample from the posterior distributions. In this section, the hybrid algorithm uses a switching threshold of \(s=10/8=1.25\).

\subsubsection*{Changing the reproductive number}
We fix \(T=25\) and perform inference for $R_0$ using the three different time series while holding all other parameters fixed at the values stated at the start of Section~\ref{SEC::SE8I8R Branching Process}. For this experiment, there is very little difference between the posterior distributions of the SE8I8R model and the SEIR model, so the plots of the posterior distributions are reserved for Appendix~\ref{APP::SE8I8R plots}. The similarity in results between the SEIR and SE8I8R models suggests that the inference problem is not sensitive the the choice of the number of E and I stages, and the approximations are not degraded significantly by increasing the number of stages.

Table~\ref{TAB::Ex2 Vary R0 SEIR} shows that the ESS for the PF is degraded significantly as \(R_0\) increases, while the ESS for the Gaussian approximation is unaffected. Moreover, comparing Table~\ref{TAB::Ex2 Vary R0 SEIR} with Table~\ref{TAB::Vary R0 SEIR}, we see that the the ESS of the PF is degraded significantly when applied to the larger SE8I8R model compared to the smaller SEIR model, while the ESS for the Gaussian approximation is similar for both models. Table~\ref{TAB::Ex2 Vary R0 SEIR} also shows that the full Gaussian approximation is slower than the particle filter for the \(R_0=1.12\) case, but is faster in the \(R_0=2.8\) case. 

\begin{table}
    \centering 
    \begin{tabular}{ccrrr}$R_0$ & Method & ESS & ESS/sec & Time (s)\\ \hline \hline 
1.12 & Gaussian & 12841.2 & 7.3 & 1768.8 \\ 
1.12 & Hybrid & 10876.2 & 9.8 & 1109.6 \\ 
1.12 & Particle & 11156.5 & 10.3 & 1086.0 \\ \hline 
2.8 & Gaussian & 12762.3 & 5.7 & 2245.3 \\ 
2.8 & Hybrid & 8813.5 & 3.0 & 2976.7 \\ 
2.8 & Particle & 7136.7 & 1.9 & 3795.4 \\ \hline 
$4.\overline 6$ & Gaussian & 13097.1 & 6.8 & 1918.0 \\ 
$4.\overline 6$ & Hybrid & 9315.3 & 4.0 & 2349.1 \\ 
$4.\overline 6$ & Particle & 1589.4 & 0.1 & 15380.2 \\ \hline 

    \end{tabular}
    \caption{Table of effective sample sizes and run times for the three different values of the \(R_0=1.12,\, 2.8,\, 4.\overline{6}\), using the three methods to evaluate the likelihood for the SE8I8R model. Run time, ESS and ESS/sec are calculated excluding burn in.}
    \label{TAB::Ex2 Vary R0 SEIR}
\end{table}

\subsubsection*{Changing the amount of data}
Now, fix \(R_0=2.8\) and let \(T=10,\,15,\,25\) vary, while holding all other parameters fixed at their true values. The posterior distributions for this case are left to Appendix~\ref{APP::SE8I8R plots} because they are similar to those in Figure~\ref{FIG:Vary R0 and T SEIR}~(right). 

Table~\ref{TAB::Ex2 Vary T SEIR} shows that the ESS of the PF is lower for larger values of \(T\), while the ESS of the Gaussian approximation is unaffected by \(T\). The run time of the Gaussian approximation is longer than the PF for small amounts of data \(T=10,15\), but is faster for larger amounts of data, \(T=25\). The most expensive calculation in the likelihood with the Gaussian approximation is the calculation of the variance. This needs to be done only once for a given set of parameters, then can be used for all time steps. Hence, as \(T\) increases, the time-cost of the expensive variance calculation can be amortised over more time-steps.

\begin{table}
    \centering 
    \begin{tabular}{ccrrr}$T$ & Method & ESS & ESS/sec & Time (s)\\ \hline \hline 
10 & Gaussian & 12884.9 & 13.8 & 936.1 \\ 
10 & Hybrid & 12538.2 & 36.2 & 346.7 \\ 
10 & Particle & 12560.7 & 36.2 & 346.7 \\ \hline 
15 & Gaussian & 12884.0 & 6.3 & 2037.2 \\ 
15 & Hybrid & 10659.6 & 6.3 & 1697.0 \\ 
15 & Particle & 11237.9 & 8.2 & 1365.2 \\ \hline 
25 & Gaussian & 12762.3 & 5.7 & 2245.3 \\ 
25 & Hybrid & 8813.5 & 3.0 & 2976.7 \\ 
25 & Particle & 7136.7 & 1.9 & 3795.4 \\ \hline 

    \end{tabular}
    \caption{Table of effective sample sizes and run times for the three different values of the \(T=30,60,90\), using the three methods to evaluate the likelihood for the SE8I8R model. Run time, ESS and ESS/sec are calculated excluding burn in.}
    \label{TAB::Ex2 Vary T SEIR}
\end{table}

\subsubsection*{Posterior distribution of the state of the branching process}
\begin{figure}
    \begin{minipage}{0.5\textwidth}
    \centering
    Exposed \(E_{4,t}\) (State 4)
    \includegraphics[width=\textwidth]{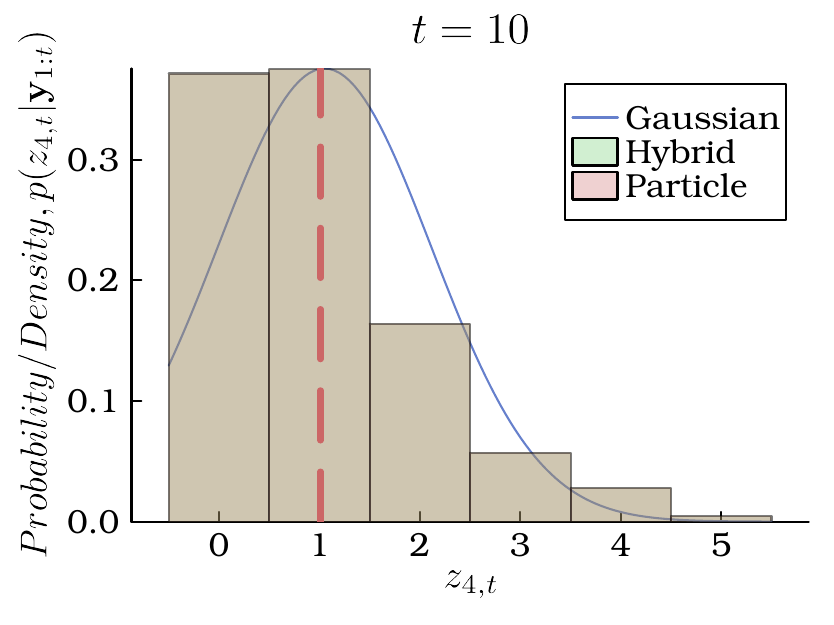}
    \includegraphics[width=\textwidth]{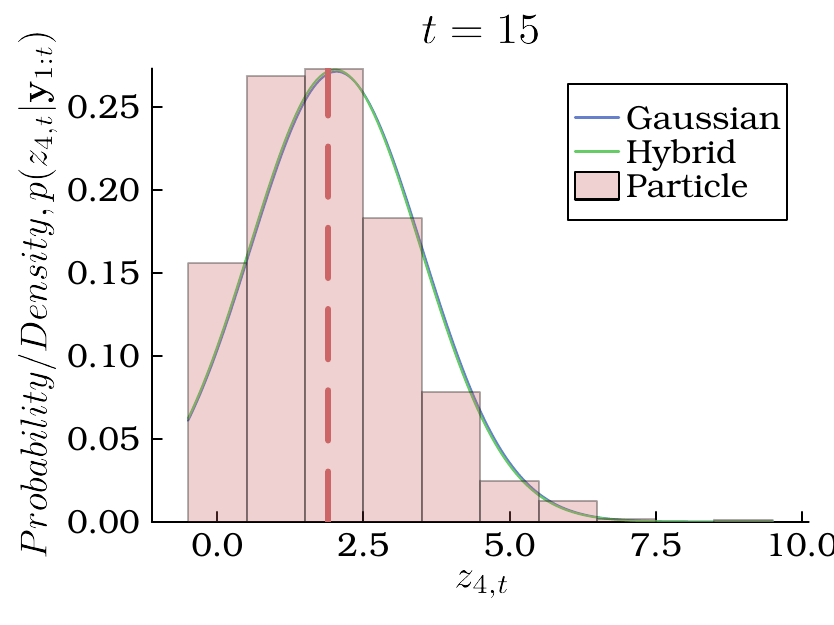}
    \includegraphics[width=\textwidth]{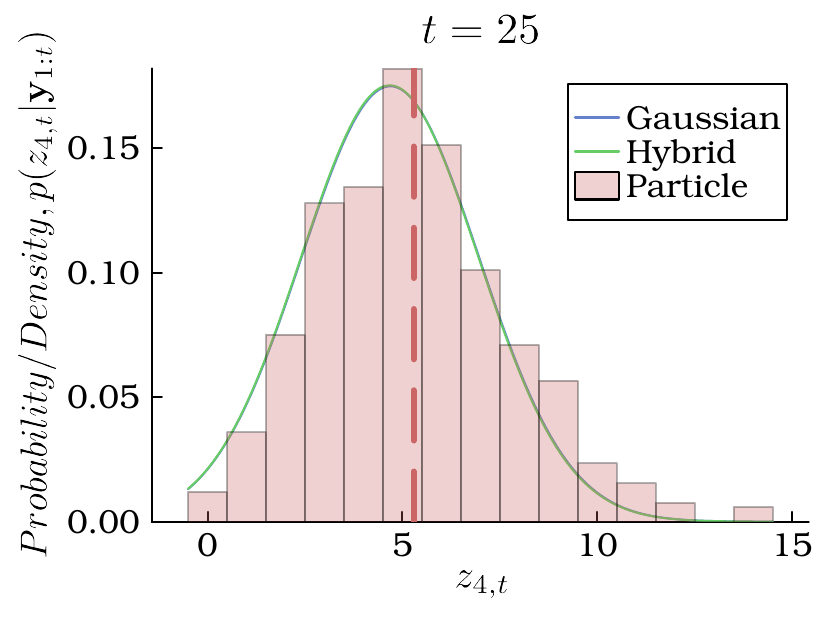}
    \end{minipage}%
    \begin{minipage}{0.5\textwidth}
    \centering
    Infectious \(I_{4,t}\) (State 12)
    \includegraphics[width=\textwidth]{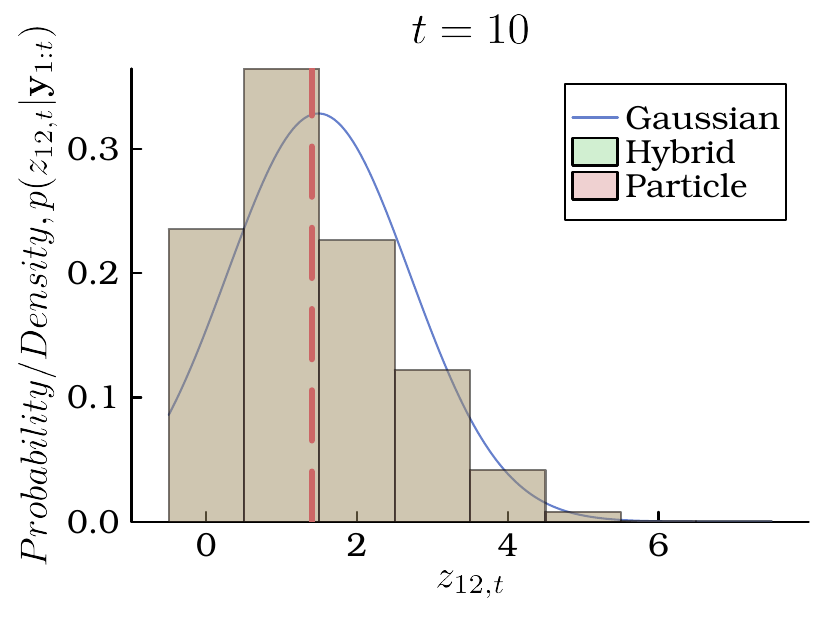}
    \includegraphics[width=\textwidth]{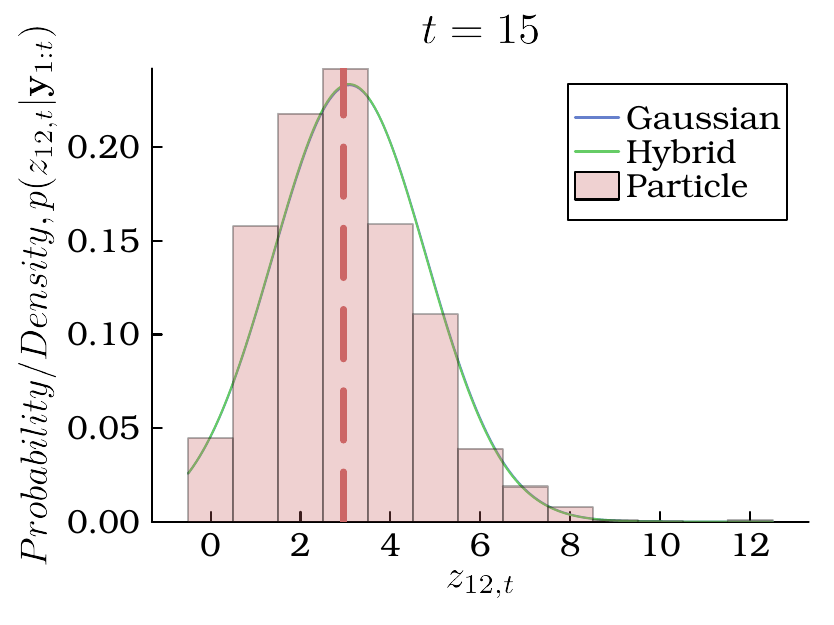}
    \includegraphics[width=\textwidth]{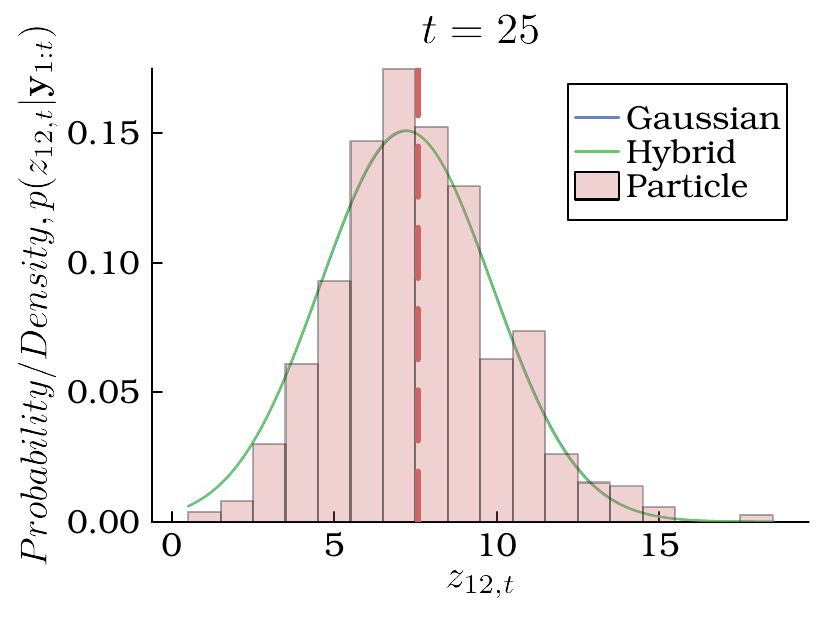}
    \end{minipage}
    \caption{\label{FIG:Ex2 state posteriors SEIR} Posterior distributions of the counts of Exposed individuals in stage 4 \(E_{4,t}\) (State 4 of the branching process) (left) and Infectious individuals in stage 4 \(I_{4,t}\) (State 12 of the branching process) (right) individuals at times \(T=10,15,25\) (top, middle, bottom, respectively) for the SE8I8R model. Distributions approximated by particles are represented as histograms, else they are represented as densities. The vertical line is at the mean of the particles from the particle filter and estimates the true posterior mean.}
\end{figure}

The posterior distributions of the state of the branching process \(p(\bs z_{t}\cond \bs y_{1:t})\) with the parameters fixed at their true values and \(R_0=2.8\) are shown in Figure~\ref{FIG:Ex2 state posteriors SEIR}. At time \(t=10\) the posterior distributions produced by the hybrid and particle methods are the same because the switching threshold has not-yet been crossed, and at times \(t=15\) and \(t=25\) the posterior distributions produced via the hybrid and particle methods differ. Figure~\ref{FIG:Ex2 state posteriors SEIR} demonstrates that the posterior distributions produced by the approximation cannot capture the truncation of the distribution at low values of \(T=10\), but the posterior distributions are reasonable at larger values of \(T=25\).

\subsection{Summary of numerical experiments}
The numerical experiments demonstrate that the Gaussian approximation is most effective in cases where case-counts are high (large values of \(R_0\) and large values of \(T\)). In these cases, the numerical experiments demonstrate that the approximations to the posterior distributions of \(R_0\) and state of the branching process are accurate, and significantly faster than the PF. Moreover, the Gaussian approximation does not suffer the some of the pit falls of the PF such as decreasing ESS and increasing run time with increasing population sizes, length of data, and number of types in the model.

\subsection{Victoria Second COVID-19 Wave }
\label{SEC::Victoria Second COVID-19 Wave}
Here we apply the Gaussian approximation to a more complex and computationally challenging problem. We use 98 days (14 weeks) of data from the second COVID-19 wave in Victoria, Australia \citep{departmentofhealthVictorianCOVID19Data2022}. During this wave daily case counts exceed 600 people for which evaluating the likelihood with the particle filter is prohibitively slow. The length of the time series means that the methods which depend on particle filtering (the particle filter and hybrid methods), will produce high-variance loglikelihood estimates, even if low switching thresholds are used. Hence, it is only feasible to use the Gaussian approximation in this case.

The primary purpose of this section is not to give novel insight into the basic reproduction number of COVID in Victoria at this time, but rather to illustrate that this method allows us to fit a relatively complex branching process model to messy data. Given the computational challenges associated with this type of estimation, it would be infeasible with a particle filter approach. 

For the branching process model we suppose there is a time-dependant reproduction number changing every 7 days \citep{nouvelletSimpleApproachMeasure2018, nishiuraEffectiveReproductionNumber2009}. Let \(R_n\) be the reproduction number for the time period \(t\in[7(n-1),7n),
, n=1,...,14\). Each $R_n$ is a parameter to be estimated from the data. The prior distribution of \(\log(R_n),\,n=1,...,14,\) is a Gaussian Process with an exponential covariance function which has mean 0, variance parameter \(\sigma_{gp}^2=0.7^2\) and length scale \(\ell=136.47\) chosen so that the correlation between \(\log(R_n)\) and \(\log(R_{n-1})\) is approximately 0.95. We use the same SEIR model as in the previous experiment (with one Exposed state and one Infectious state). The mean latent period (mean time of individuals in the exposed state) is fixed at \(T_E=2.0=1/\delta\), the mean infectious period at \(T_I=1.0=1/\lambda\), the observation probability is \(p=0.75\) and the observation variance is \(\sigma=20\). The rate at which an individual creates new infections at time \(t\in[7(n-1),7n)\) is \(\beta_n=\lambda R_n,\,n=1,...,14\). A short infectious period was chosen to reflect that, during this period, most individuals would isolate quickly after symptoms or testing.
We also infer the initial number of exposed and infective individuals at time \(t=0\). The prior distribution on the initial number of exposed and infective individuals is multivariate Gaussian with a mean vector \( [10, 10]^T\) and variance matrix \(10\bs I\).

\begin{figure}
    \centering
    \includegraphics[width=\textwidth]{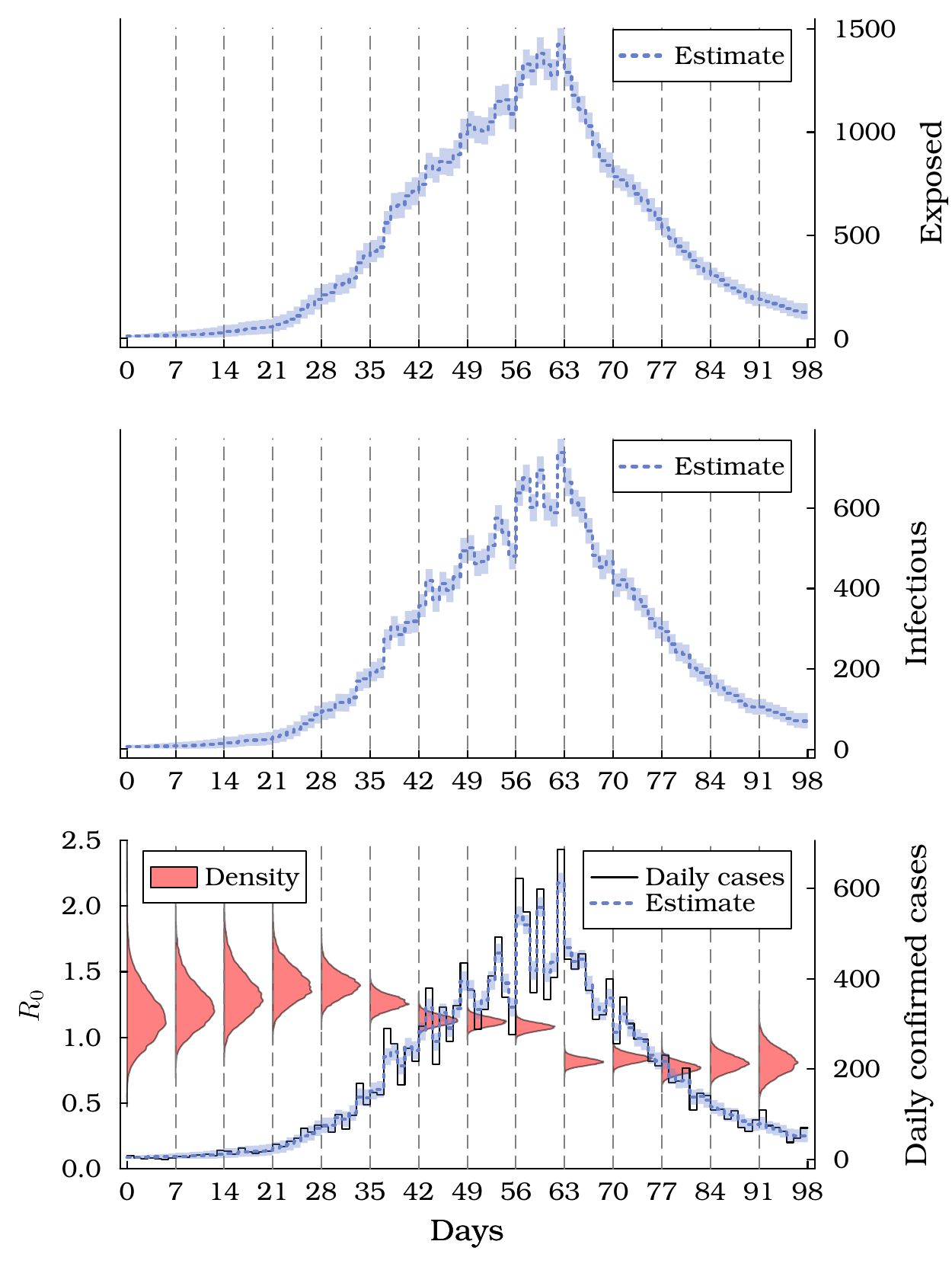}
    \caption{Top: median and symmetric 80\% credible interval (80\% CI) of the filtering distribution for exposed individuals, \(p(z_{t,1}\cond \mathbf y_{1:t})\). Middle: median and 80\% CI of the filtering distribution for infectious individuals, \(p(z_{t,2}\cond \mathbf y_{1:t})\). Bottom: half-violin plots showing the posterior densities for time-varying reproduction numbers, \(R_n,\,n=1,...,14\), as well as the daily case counts from the second COVID-19 wave in Victoria and the median and 80\% CI of the filtering distribution for new daily cases, \(p(z_{t,3}\cond \mathbf y_{1:t})\).}
    \label{FIG::Ex3 Marginal1}
\end{figure}

The bottom plot in Figure~\ref{FIG::Ex3 Marginal1} shows the posterior distributions for each of the time-varying reproduction numbers and demonstrates that the posterior distributions appear to be reasonable estimates given the data. In particular, we see that the estimates for $R_n$ drop below 1 as the daily case counts begin to fall. Additionally, we see that earlier $R_n$ estimates have far more variance than later estimates. We believe that this occurs since there are fewer infectious individuals at at earlier times, meaning there are fewer infection events and therefore less information about the infection rate of the disease. Towards the end of the time series, we observe a slight increase in the uncertainty in the posterior distributions of $R_n$ as case numbers are again lower so there is less data to inform the infection rate. 

The top, middle and bottom plots in Figure~\ref{FIG::Ex3 Marginal1} show the estimated medians and symmetric 80\% credible intervals of the filtering distributions \(p(z_{t,i}\cond \mathbf y_{1:t}),\, i=1,2,3,\) for the number of exposed individuals, infectious individuals and observed new daily cases, respectively. Intuitively, Figure~\ref{FIG::Ex3 Marginal1} shows that the number of infectious individuals is approximately \(1/p=4/3\) times the number of confirmed daily cases, and the number of exposed individuals is approximately \(T_E/T_I=2\) times the number of infectious individuals. Figure~\ref{FIG::Ex3 Marginal1} also shows some evidence that the trajectories of the filtered state estimate is smoother for exposed individuals than for infectious individuals, which is then smoother than the filtered state estimate of the daily new cases.

\begin{figure}
    \centering
    \includegraphics[width=\textwidth]{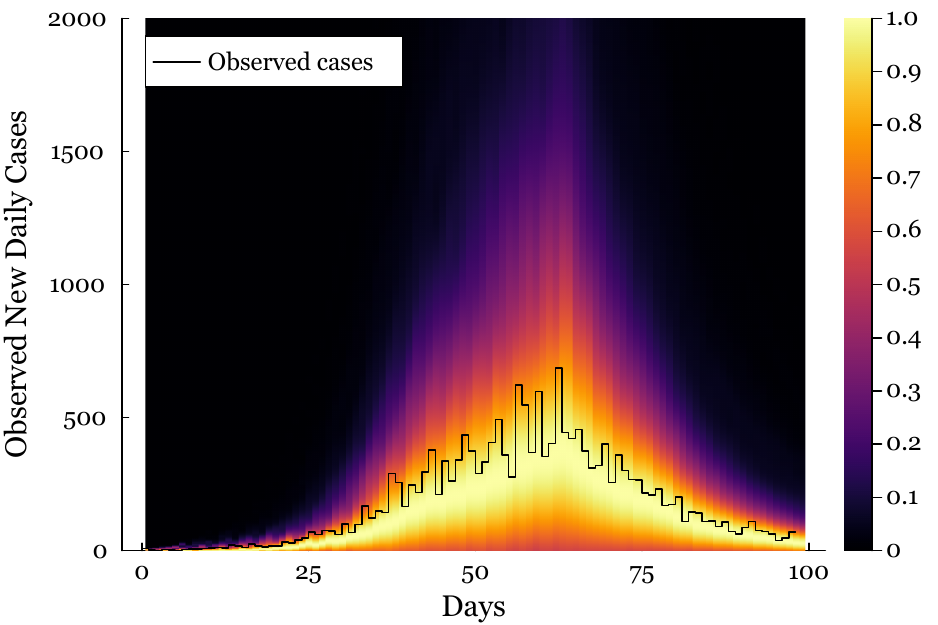}
    \caption{Heat-map of simulated trajectories of the SEIR branching process with parameters randomly sampled from the posterior distribution for second COVID-19 wave in Victoria. The observed data is plotted as a black line.}
    \label{FIG::trajectories}
\end{figure}

Figure~\ref{FIG::trajectories} shows a posterior predictive plot of trajectories of the branching process simulated with parameters from the posterior distribution. The observed data lies in a high-density region of the posterior predictive distribution, suggesting that the model is reasonable. 

The procedure took $96.5$ seconds to complete on a single thread\footnote{Code was written in Julia v1.11 and run on a Macbook Air with M2 processor.}. The total number of samples was 831,072, and the minimum ESS was 6,469, after removing 131,072 burn-in samples. The convergence of the chain was validated with the $\hat{R}$ diagnostic \citep{brooksHandbookMarkovChain2011}. 



\section{Conclusion}





While multitype continuous-time branching processes can be useful for modelling, in the cases where agent counts become large, state and parameter estimation using a particle filter can become infeasible. In this paper we have presented an algorithm for approximate state and likelihood estimation that can perform significantly faster than a standard particle filter, making it more feasible to be used as part of a Metropolis-Hastings scheme for parameter estimation \citep{sarkkaBayesianFilteringSmoothing2013}. The likelihood estimates produced via this approximation are not estimates with an intrinsic variance, which also improves the performance of the Metropolis-Hastings sampler. We have demonstrated through simulation studies how our proposed approximation behaves for both the SEIR and SE8I8R branching process models. 

In general, our approximation tends to overestimate the states of the system when the populations are smaller and hence there is a larger probability that the population will be zero (see Figure \ref{FIG:state posteriors SEIR}). In our context of epidemic modelling, this overestimation of the number of infected then leads to an underestimation of $R_0$ (see Figure \ref{FIG:Vary R0 and T SEIR}) as lower growth is required to produce the same case numbers data.
One direction for further study is to attempt to reduce the bias in the posterior distributions caused by considering only the first two moments. Two possible avenues to consider are to either use higher order moments in the approximation, beyond the mean and variance \citep{sophie,thomas:2015}, or to correct for the bias introduced by the approximation \citep{bon2025}.

We have demonstrated that the approximation is accurate when agent counts are high, which is particularly useful because particle filter methods can be prohibitively slow in these cases. When agent counts are are smaller, we propose a switching method which makes strategic use of particle filters when agent counts are low, and uses our approximation when case counts are high, to minimise both computation time and bias. We have demonstrated the accuracy and applicability of this approach, which decreases bias, but at an increased computational expense.

For longer time series, using the approximation for all times is feasible as demonstrated by our COVID-19 example. Here we chose to infer the weekly, rather than daily, transmission rate. This is motivated not to save on computational effort, but because there are strong day-of-the-week effects in this data that we do not attempt to capture in the model. Thus by keeping the transmission fixed over a week we can partially mitigate this model misspecification. The inclusion of the observation noise as a regularisation also help obtain a better fit. Testing with smaller values of $\sigma$ produced much worse estimates of the transmission rates as evaluated by posterior predictive checks similar to that shown in Figure \ref{FIG::trajectories}.
Our formulation of the observation model assumes that the regularisation noise is constant over time. Our results here could be improved by making this a function of the data as there is clearly higher noise when the observed cases counts are large around the peak of the epidemic.

Given the general formulation of the Gaussian approximation, it could theoretically be used for any continuous-time branching process. The main bottleneck in increasing the complexity of the model is the dimension of the state space. In particular, the addition of more intermediary states will mean that the number of agents will be spread out across many agent types. For the particle filter, this means that the overall number of events will be higher and the particle filter will take even longer to run. For the Gaussian approximation, the calculation of the mean and variance requires computations involving matrices of size $r(r+1)\times r(r+1)$, where $r$ is the number of types, which may be problematic for models with very high-dimensional state spaces. 
Future work is to reduce the computation time for the variance matrix. Using the same techniques as those applied to Markovian Binary Trees \citep{sophie}, it can be shown that the variance matrix satisfies a Sylvester equation which can be solved more efficiently than the matrix exponential calculation used here. However, this equation does not always have a unique solution and we do not know which solution to the Sylvester equation we require. In particular, the SEIR models presented in this work results in a Sylvester equation which does not have a unique solution for any values of the parameters.





\section*{Appendix}
\appendix
\section{Further results of numerical experiments}\label{APP: further sims}
This appendix contains further results on numerical experiments to demonstrate the performance of the approximation scheme.
\subsection{SEIR model}
\subsubsection*{Changing the observation variance}
\begin{figure}
    \begin{minipage}{0.5\textwidth}
    \centering
    Vary \(\sigma^2\)
    \includegraphics[width=\textwidth]{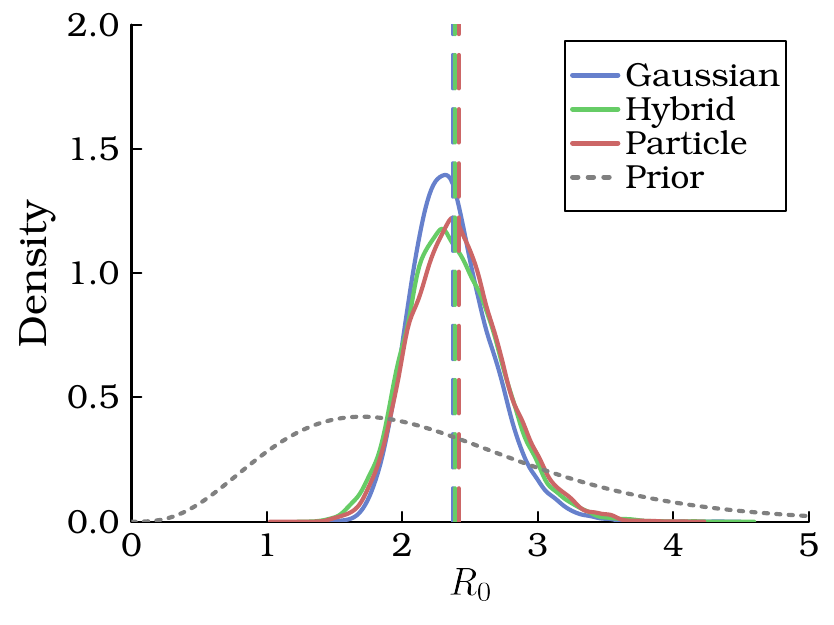}
    \includegraphics[width=\textwidth]{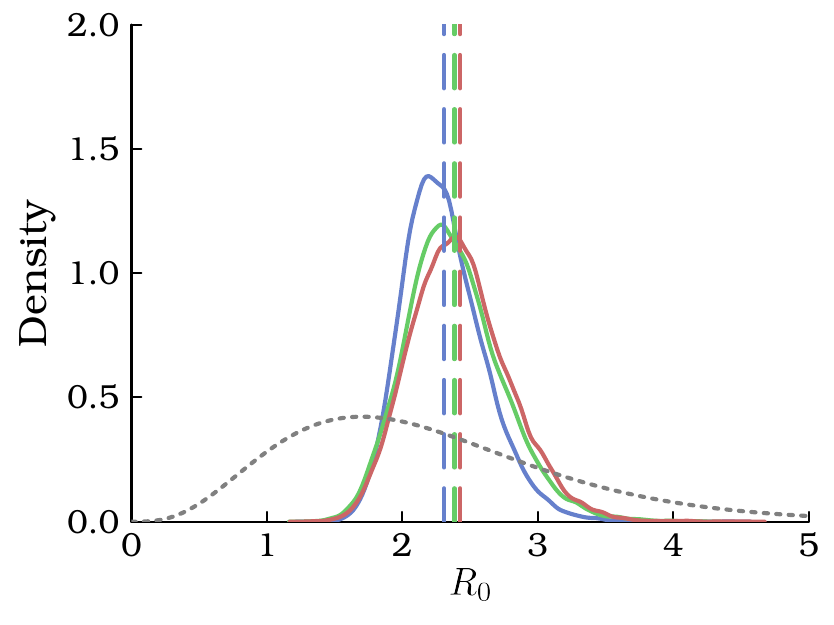}
    \includegraphics[width=\textwidth]{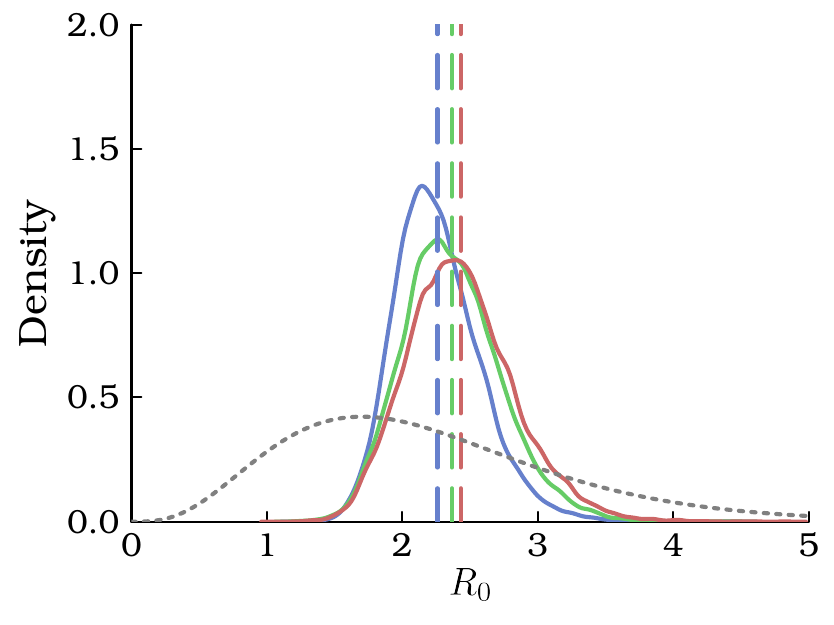}
    \end{minipage}%
    \begin{minipage}{0.5\textwidth}
    \centering
    Vary \(p\)
    \includegraphics[width=\textwidth]{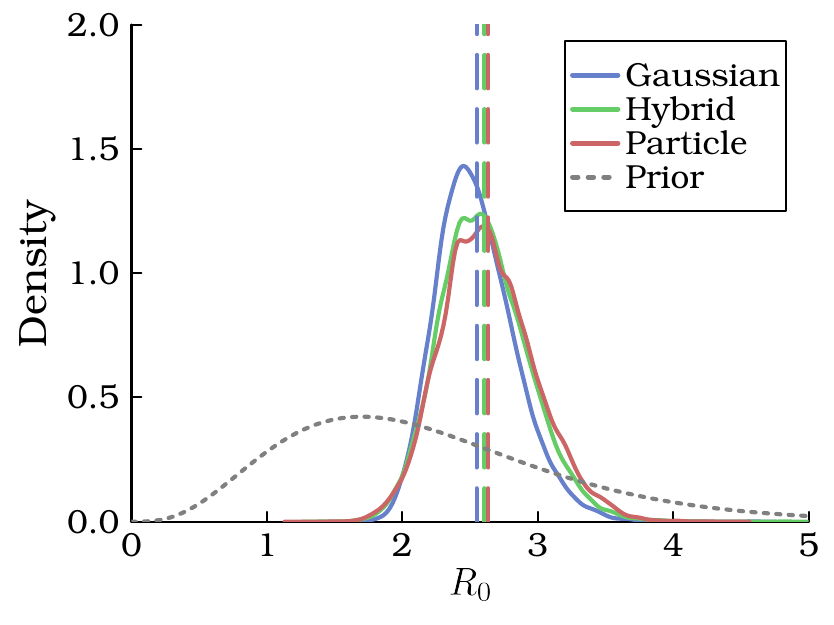}
    \includegraphics[width=\textwidth]{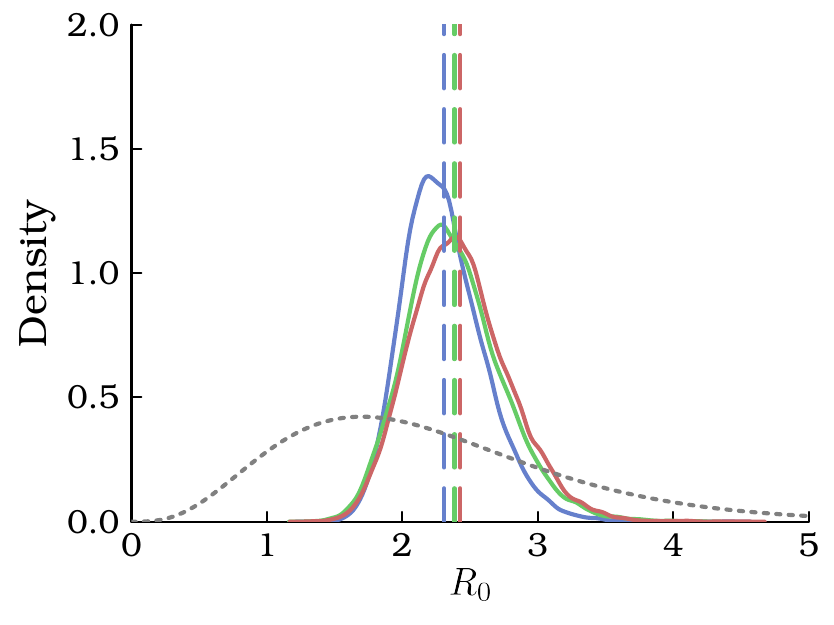}
    \includegraphics[width=\textwidth]{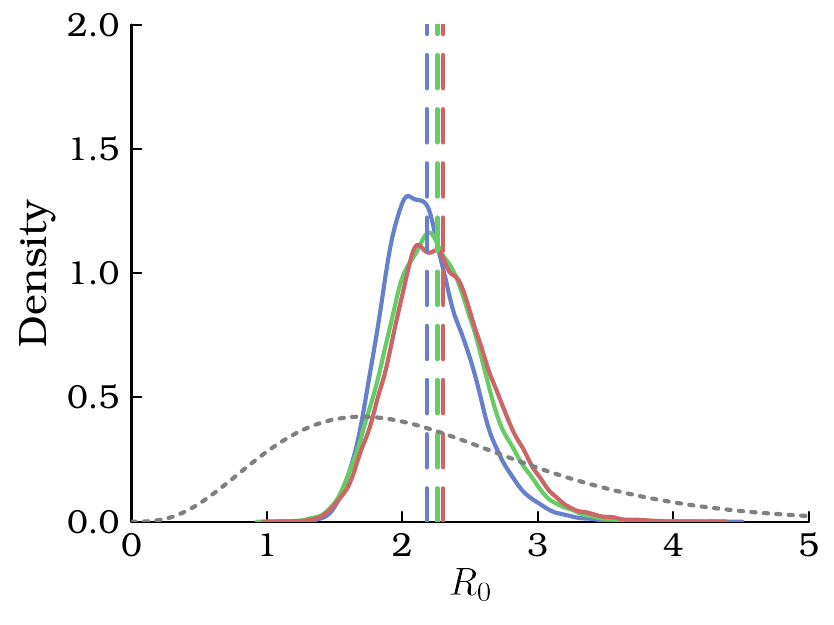}
    \end{minipage}
    \caption{\label{FIG:Vary S and p SEIR}(Left) Posterior distributions for three different values of the observation variance (regularisation parameter) \(\sigma^2=0.25,\,1,\,4\), (top, middle and bottom, respectively) using the three methods to evaluate the likelihood for the SEIR model. (Right) Posterior distributions for three different values of the observation probability \(p=0.5,\,0.75,\,1\), (top, middle and bottom, respectively) using the three methods to evaluate the likelihood for the SEIR model. The vertical lines are at the mean of the posterior distributions.}
\end{figure}

Here we use the \(R_0=2.8\) and \(T=25\) simulated dataset and vary the observation variance \(\sigma^2=0.25,\,1,\,4\), while holding all other parameters fixed at their true values. Recall that the true value of \(\sigma^2=0\), and we can think of \(\sigma^2\) as a regularisation parameter, with larger values providing more regularisation. Regularisation can be used to prevent over-fitting, and alleviate issues with model misspecification. Figure~\ref{FIG:Vary S and p SEIR}~(left) shows the posterior distributions for using the three methods for the three different values of \(\sigma^2\).

Figure~\ref{FIG:Vary S and p SEIR}~(left) shows that the mean of true posterior distribution (produced using the particle filter) does not appear to change as a function of the regularisation parameter, but the variance of the posterior distribution increases slightly. Figure~\ref{FIG:Vary S and p SEIR}~(left) shows a small amount of evidence that the Gaussian approximation gets worse as the amount of regularisation decreases (\(\sigma^2\) gets larger). 

In Table~\ref{TAB:Vary S SEIR} we see that the ESS for the Gaussian approximation is unaffected by the value of \(\sigma^2\), while the ESS for the hybrid and particle methods reduces for small values of \(\sigma^2=0.25\). Intuitively, a larger value of \(\sigma^2\) permits a greater difference between the observed data and the simulated particles (trajectories) in the particle filter, increasing the diversity of the particles. We hypothesize that this decreases the variance of the likelihood estimates, which improves the sampler resulting in a higher ESS.

\begin{table}
    \centering 
    \begin{tabular}{ccrrr}$\sigma^2$ & Method & ESS & ESS/sec & Time (s)\\ \hline \hline 
0.25 & Gaussian & 13267.5 & 6907.2 & 1.9 \\ 
0.25 & Hybrid & 7215.4 & 30.2 & 239.3 \\ 
0.25 & Particle & 6317.9 & 11.6 & 543.9 \\ \hline 
1.0 & Gaussian & 12726.6 & 6792.1 & 1.9 \\ 
1.0 & Hybrid & 9631.1 & 42.6 & 225.8 \\ 
1.0 & Particle & 9161.7 & 16.1 & 569.9 \\ \hline 
4.0 & Gaussian & 13012.6 & 6727.4 & 1.9 \\ 
4.0 & Hybrid & 11228.7 & 46.9 & 239.4 \\ 
4.0 & Particle & 10982.4 & 19.3 & 568.5 \\ \hline 

    \end{tabular}
    \caption{Table of effective sample sizes and run times for three different values of the observation variance (regularisation parameter) \(\sigma^2=0.25,\,1,\,4\), (top, middle and bottom, respectively) using the three methods to evaluate the likelihood for the SEIR model. Run time, ESS and ESS/sec are calculated excluding burn in.}
    \label{TAB:Vary S SEIR}
\end{table}

\begin{figure}
\begin{minipage}{0.5\textwidth}
    \centering
    \(p=0.75,\, \sigma^2=0.25\)
    \includegraphics[width=\textwidth]{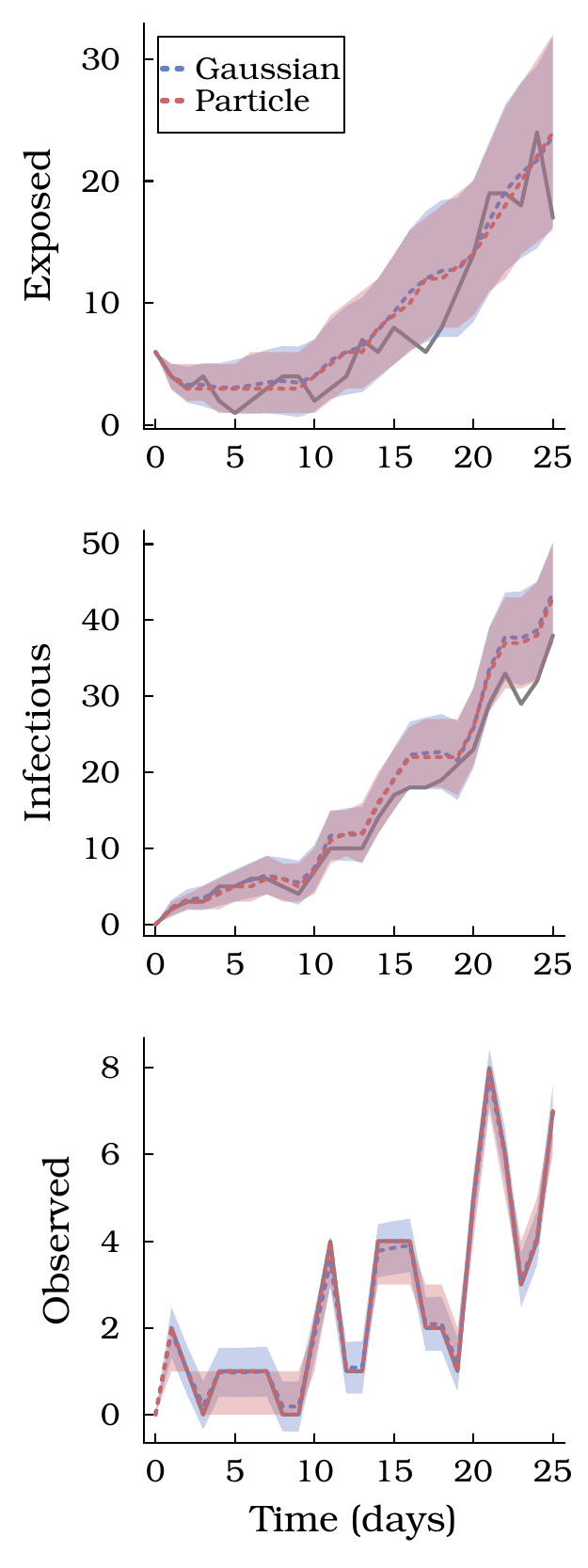}
\end{minipage}%
\begin{minipage}{0.5\textwidth}
    \centering
    \(p=0.75,\,\sigma^2=4\)
    \includegraphics[width=\textwidth]{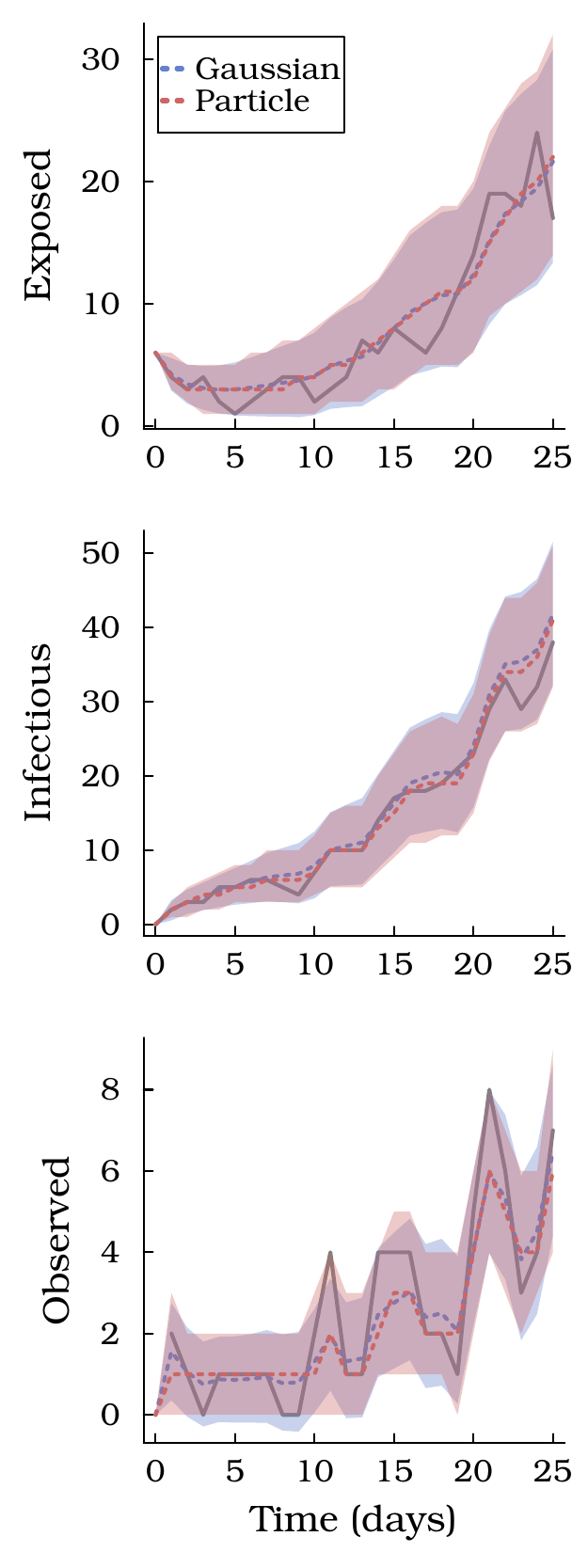}
\end{minipage}
\caption{\label{FIG:filtered var SEIR data}The median and symmetric 80\% credible interval of the filtering distribution, \(p(z_{t,i}\cond \mathbf y_{1:t}),\, i=1,2,3\), estimated by the Gaussian approximation (blue) and PF (red) for \(p=0.75,\, \sigma^2=0.25\) (left) and \(p=0.75,\,\sigma^2=4\) (right), and the simulated realisations of the SEIR epidemic model for \(R_0=\beta/\lambda=2.8\) (solid grey lines).}
\end{figure}

Figures~\ref{FIG:filtered var SEIR data} and \ref{FIG:filtered p/var SEIR data} show the medians and symmetric 80\% credible intervals of the filtering distributions, \(p(z_{t,i}\cond \mathbf y_{1:t}),\, i=1,2,3\), estimated by the Gaussian approximation and PF as \(\sigma^2\) varies. The plots demonstrate good agreement between the quantiles of the filtering distributions for the approximation and the true values (the quantiles from the particle filter). The figures also demonstrate that larger values of \(\sigma^2\) produce wider credible intervals for all types.

\subsubsection*{Changing the observation probability}
Here we use the \(R_0=2.8\) and \(T=25\) dataset and let the observation probability \(p=0.5,\,0.75,\,1\) vary, while holding all other parameters fixed at their true values. Figure~\ref{FIG:Vary S and p SEIR}~(right) shows the posterior distributions for using the three methods for the three different values of \(p\). Recall that the true value of \(p=0.75\).

Figure~\ref{FIG:Vary S and p SEIR}~(right) demonstrates that a lower value of \(p\) results in a larger value of \(R_0\) being inferred. Intuitively, this make sense; with a lower probability of observing cases, then a larger epidemic (larger \(R_0\)) is required to produce the same number of cases. 

Figure~\ref{FIG:Vary S and p SEIR}~(right) shows no strong evidence that the accuracy of the approximation is affected by changing \(p\).

Table~\ref{TAB:Vary p SEIR} shows no strong evidence that ESS depends on \(p\). Table~\ref{TAB:Vary p SEIR} does show evidence that the run time of the particle filter is negatively related to \(p\); as \(p\) gets smaller, the run time increases. This is consistent with our hypothesis above that a smaller \(p\) implies a larger epidemic for fixed data. 

\begin{table}
    \centering 
    \begin{tabular}{ccrrr}$p$ & Method & ESS & ESS/sec & Time (s)\\ \hline \hline 
0.5 & Gaussian & 12764.9 & 6775.4 & 1.9 \\ 
0.5 & Hybrid & 10045.5 & 48.8 & 205.7 \\ 
0.5 & Particle & 9132.5 & 11.7 & 779.5 \\ \hline 
0.75 & Gaussian & 12726.6 & 6792.1 & 1.9 \\ 
0.75 & Hybrid & 9631.1 & 42.6 & 225.8 \\ 
0.75 & Particle & 9161.7 & 16.1 & 569.9 \\ \hline 
1.0 & Gaussian & 12931.8 & 6894.9 & 1.9 \\ 
1.0 & Hybrid & 9471.8 & 37.2 & 254.5 \\ 
1.0 & Particle & 9057.9 & 19.5 & 464.1 \\ \hline 

    \end{tabular}
    \caption{Table of effective sample sizes and run times for three different values of the observation probability \(p=0.5,\,0.75,\,1\), (top, middle and bottom, respectively) using the three methods to evaluate the likelihood for the SEIR model. Run time, ESS and ESS/sec are calculated excluding burn in.}
    \label{TAB:Vary p SEIR}
\end{table}

\begin{figure}
\centering
\begin{minipage}{0.5\textwidth}
    \centering
    \(p=0.75,\, \sigma^2=1,\, R_0=2.8\)
    \includegraphics[width=\textwidth]{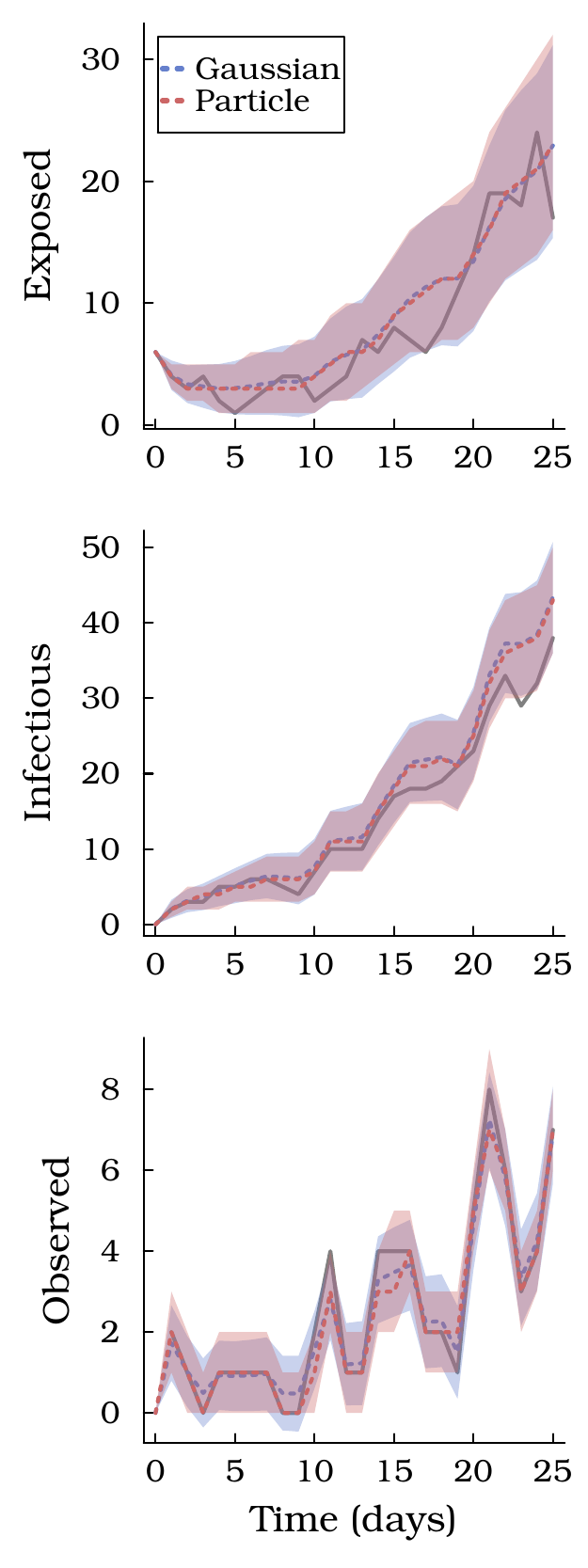}
\end{minipage}%
\caption{\label{FIG:filtered p/var SEIR data}The median and symmetric 80\% credible interval of the filtering distribution, \(p(z_{t,i}\cond \mathbf y_{1:t}),\, i=1,2,3\), estimated by the Gaussian approximation (blue) and PF (red) for \(p=0.75,\, \sigma^2=0.25\), and the simulated realisations of the SEIR epidemic model for \(R_0=\beta/\lambda=2.8\) (solid grey lines).}
\end{figure}

\begin{figure}
\begin{minipage}{0.5\textwidth}
    \centering
    \(p=0.5,\,\sigma^2=1\)
    \includegraphics[width=\textwidth]{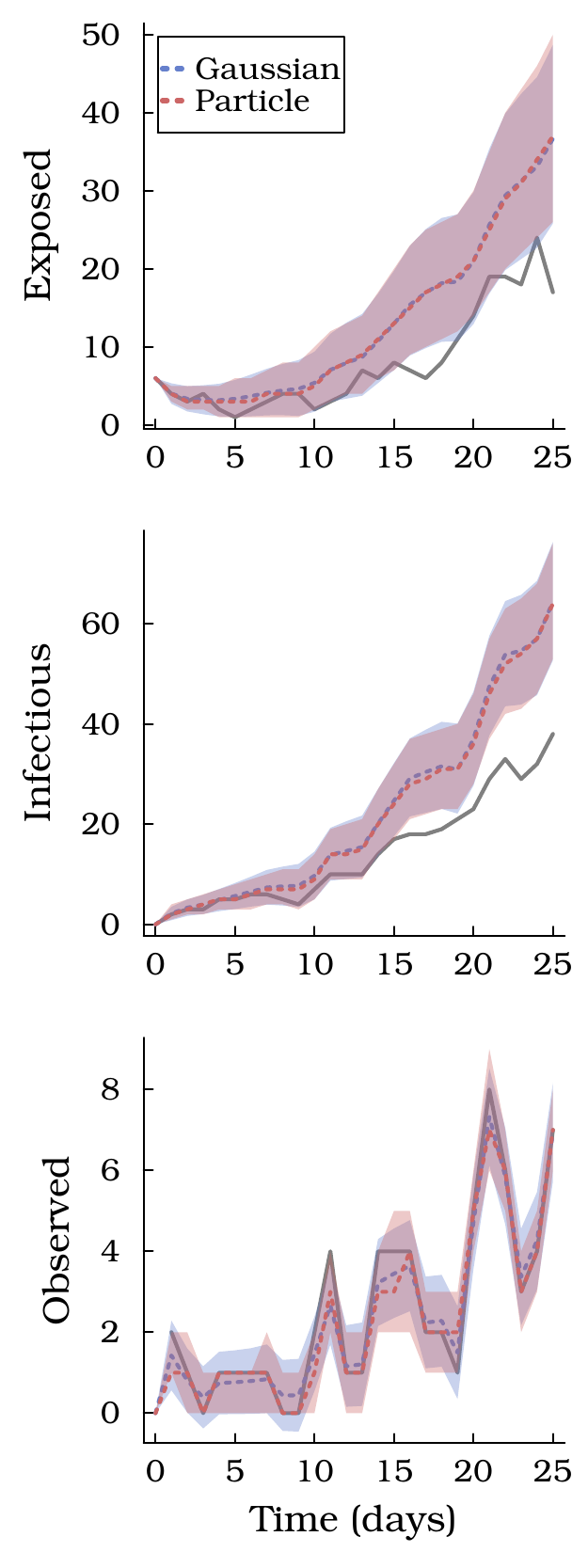}
\end{minipage}%
\begin{minipage}{0.5\textwidth}
    \centering
    \(p=1,\,\sigma^2=1\)
    \includegraphics[width=\textwidth]{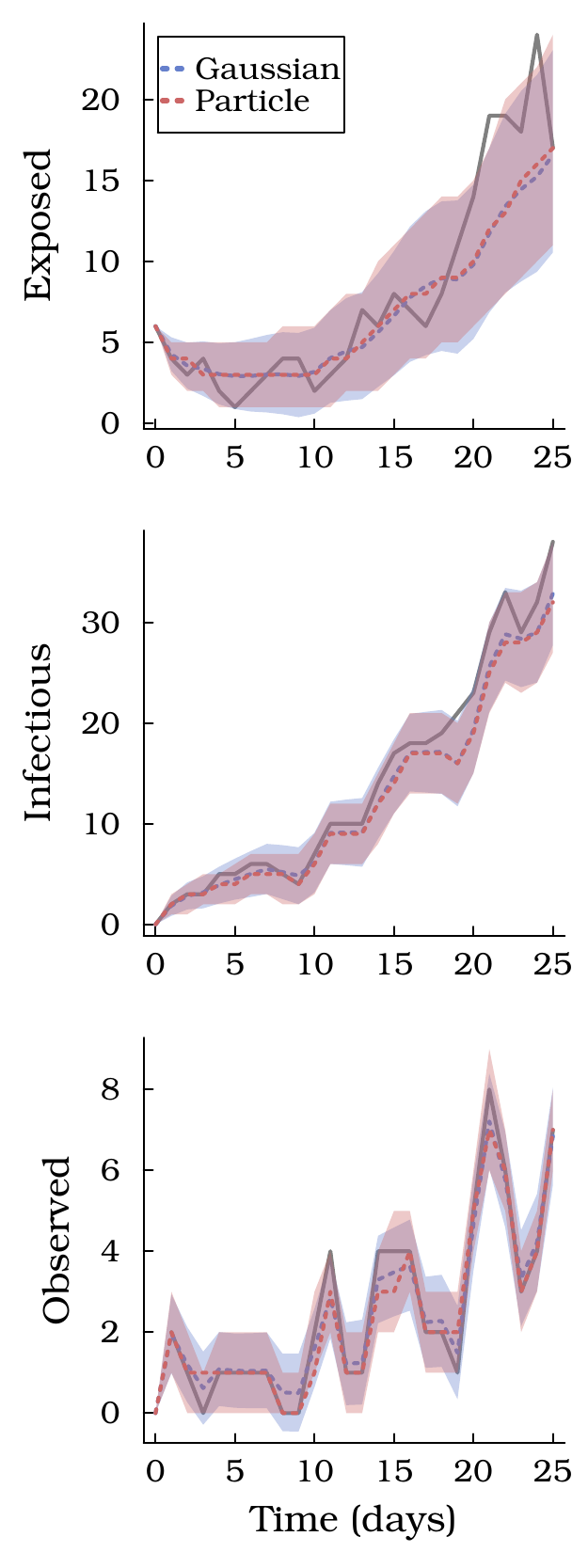}
\end{minipage}
\caption{\label{FIG:filtered p SEIR data}The median and symmetric 80\% credible interval of the filtering distribution, \(p(z_{t,i}\cond \mathbf y_{1:t}),\, i=1,2,3\), estimated by the Gaussian approximation (blue) and PF (red) for \(p=0.5,\,\sigma^2=1\) (left) and \(p=1,\,\sigma^2=1\) (right), and the simulated realisations of the SEIR epidemic model for \(R_0=\beta/\lambda=2.8\) (solid grey lines).}
\end{figure}

Figures~\ref{FIG:filtered p/var SEIR data} and \ref{FIG:filtered p SEIR data} show the medians and symmetric 80\% credible intervals of the filtering distributions, \(p(z_{t,i}\cond \mathbf y_{1:t}),\, i=1,2,3\), estimated by the Gaussian approximation and PF as \(p\) varies. The plots demonstrate good agreement between the quantiles of the filtering distributions for the approximation and the true values (the quantiles from the particle filter). The true value of \(p=0.75\) (Figure~\ref{FIG:filtered p/var SEIR data}). The figures show that when \(p=0.5\), which is less than the true value of \(p=0.75\), the number of exposed and infectious individuals is overestimated; when \(p=1\), which is greater than the true value of \(p=0.75\), the estimates of the number of exposed and infectious individuals is less when \(p=1\) compared to \(p=0.75\). However, the number of exposed and infectious individuals is still estimated reasonably well as the data lies mostly within the 80\% credible interval.

\subsection{SE8I8R model} \label{APP::SE8I8R plots}
\begin{figure}
\begin{minipage}{0.5\textwidth}
    \centering
    Vary \(R_0\)
    \includegraphics[width=\textwidth]{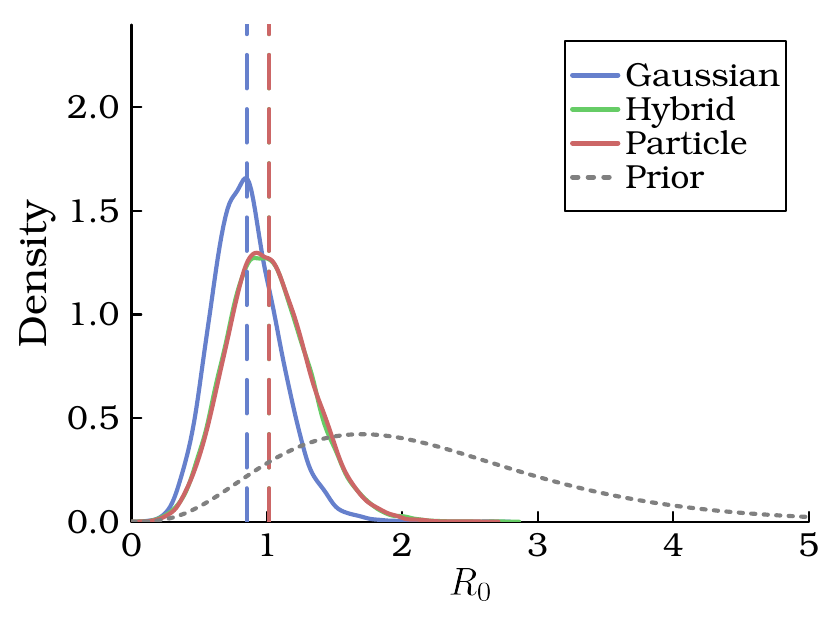}
    \includegraphics[width=\textwidth]{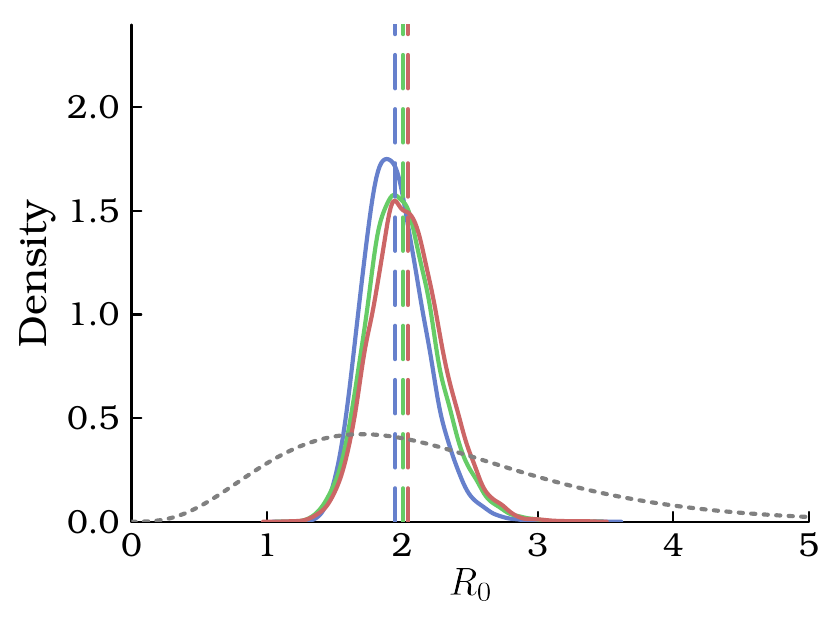}
    \includegraphics[width=\textwidth]{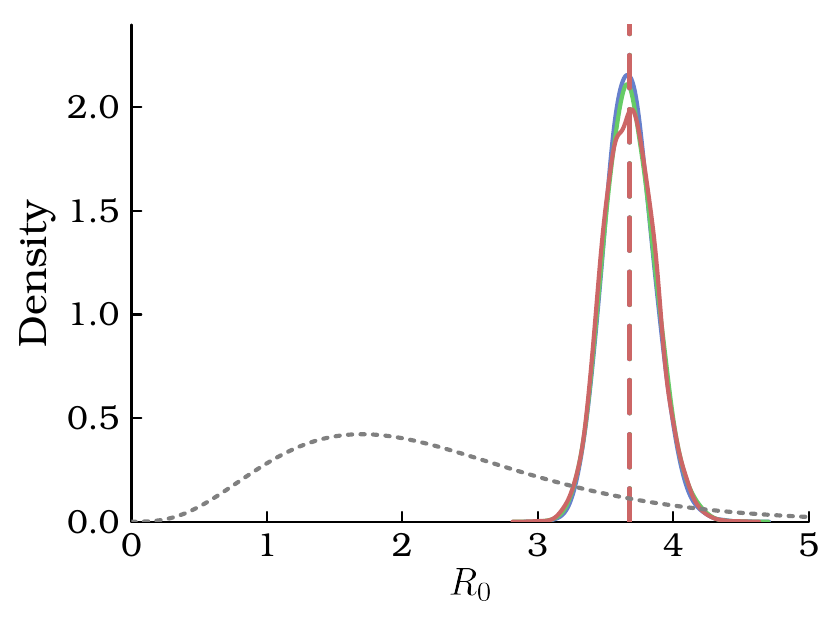}
\end{minipage}%
\begin{minipage}{0.5\textwidth}
    \centering
    Vary \(T\)
    \includegraphics[width=\textwidth]{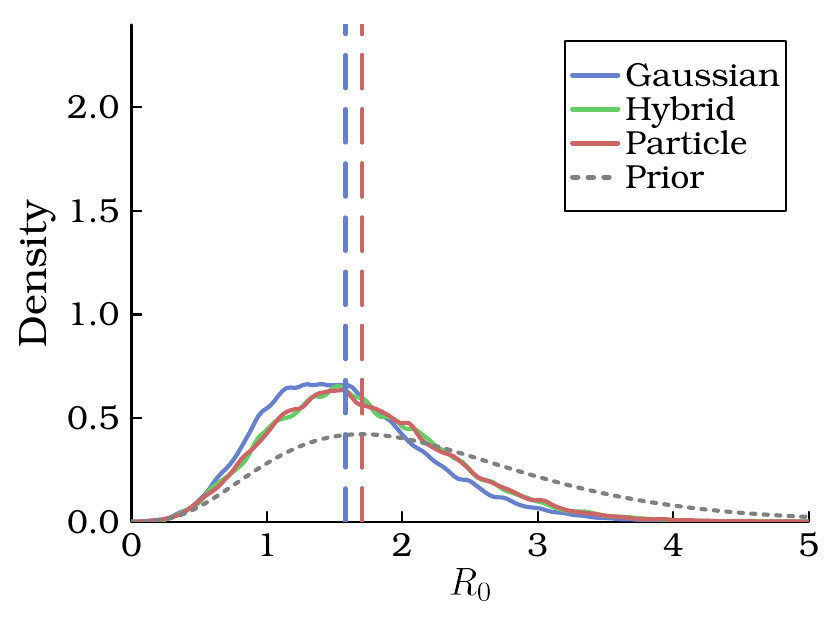}
    \includegraphics[width=\textwidth]{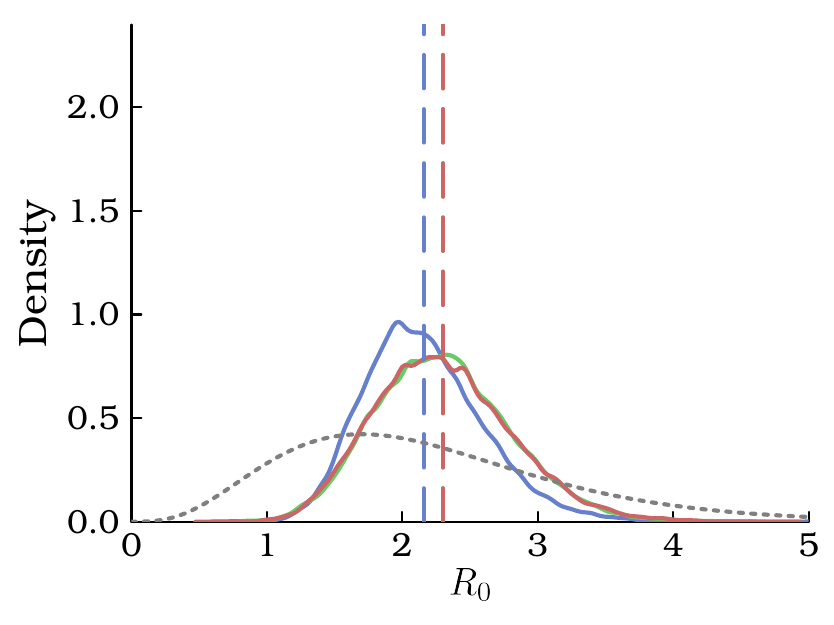}
    \includegraphics[width=\textwidth]{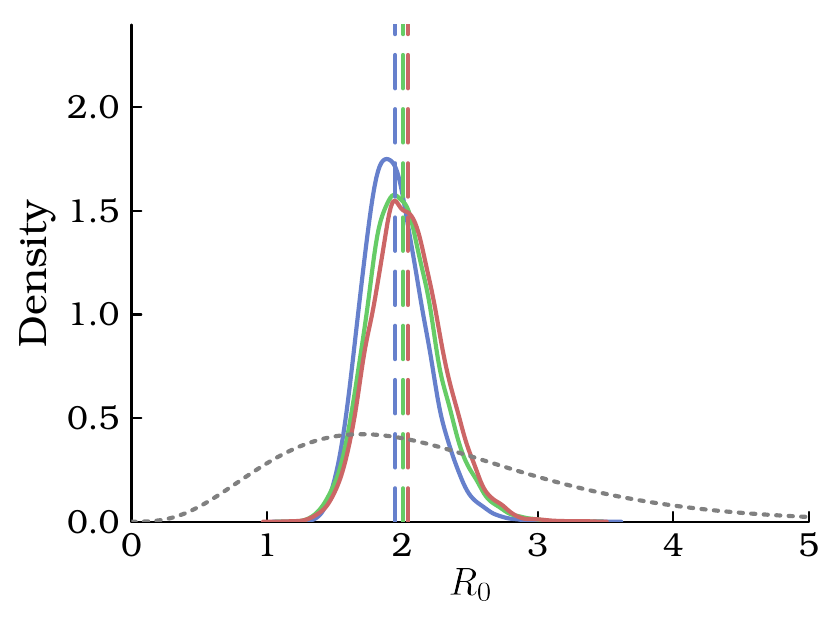}
\end{minipage}
\caption{\label{FIG:Vary R0 and T SE8I8R}(Left) Posterior distributions for three different values of the reproductive number \(R_0=\beta/\lambda=1.12,\,2.8,\,4.\overline{6}\), (top, middle and bottom, respectively) and \(T=25\) using the three methods to evaluate the likelihood for the SE8I8R model. (Right) Posterior distributions for three different lengths of time-series \(T=10,\,15,\,25\), (top, middle and bottom, respectively) and \(R_0=2.8\) using the three methods to evaluate the likelihood for the SE8I8R model. The vertical lines are at the mean of the posterior distributions.}
\end{figure}

\subsubsection*{Changing the observation variance}
\begin{figure}
    \begin{minipage}{0.5\textwidth}
    \centering
    Vary \(\sigma^2\)
    \includegraphics[width=\textwidth]{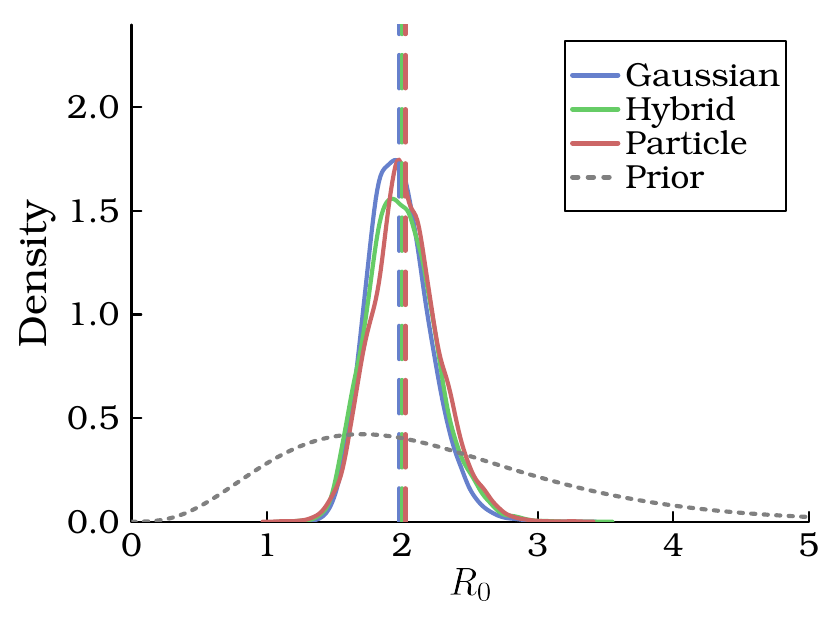}
    \includegraphics[width=\textwidth]{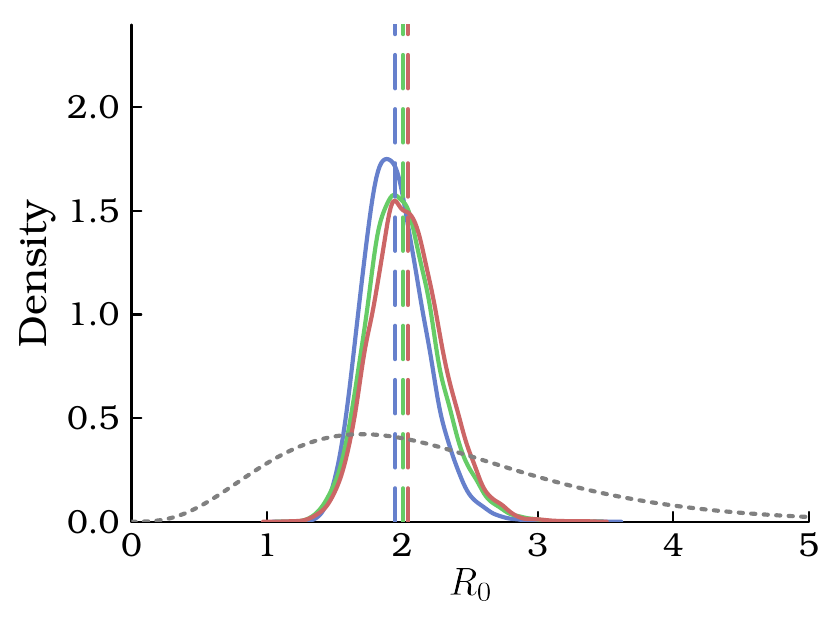}
    \includegraphics[width=\textwidth]{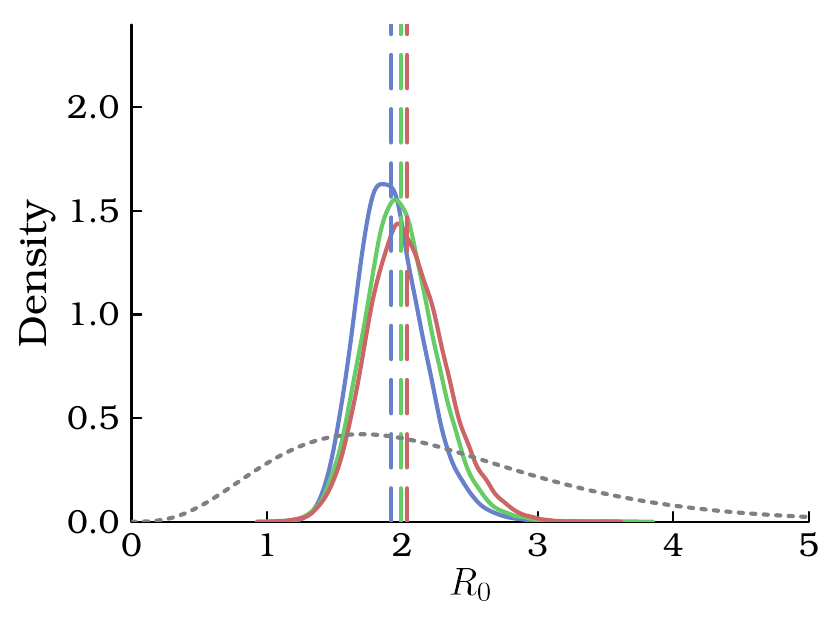}
    \end{minipage}%
    \begin{minipage}{0.5\textwidth}
    \centering
    Vary \(p\)
    \includegraphics[width=\textwidth]{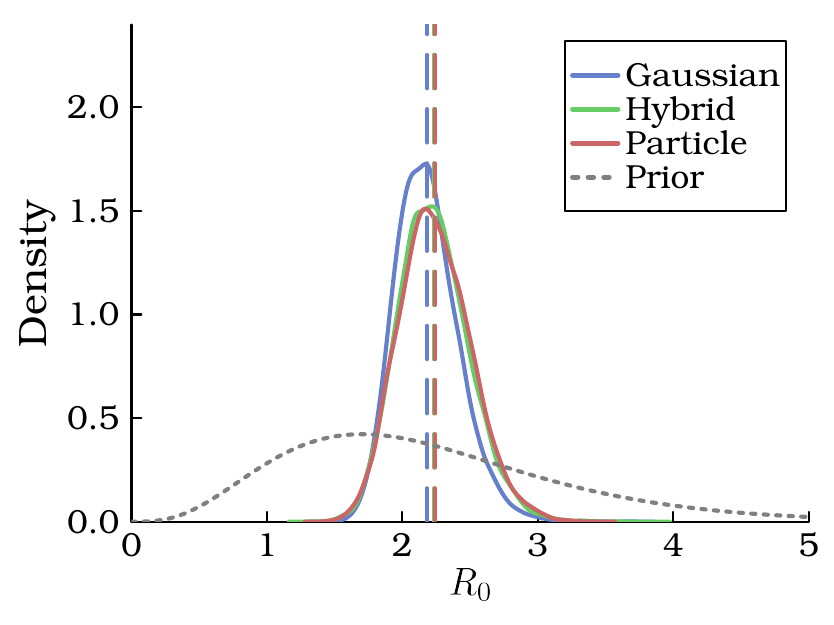}
    \includegraphics[width=\textwidth]{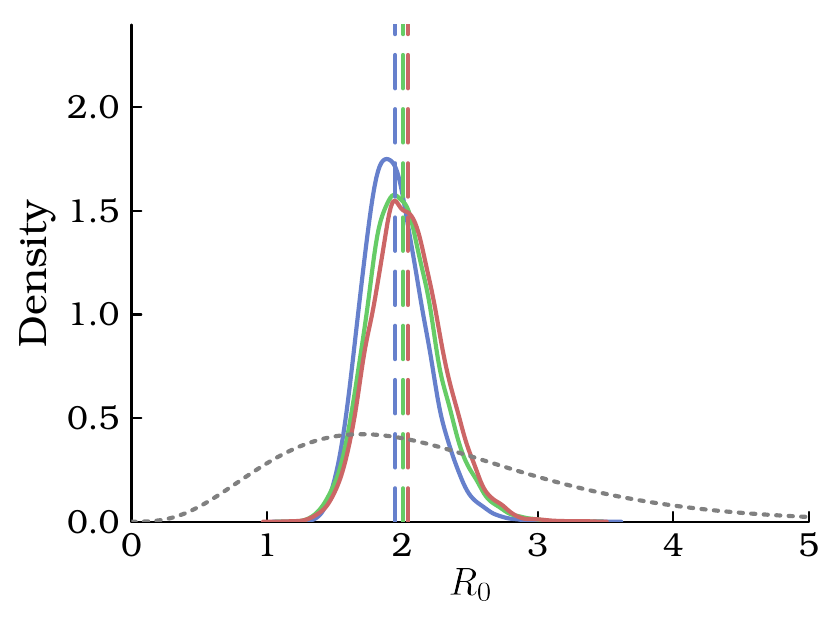}
    \includegraphics[width=\textwidth]{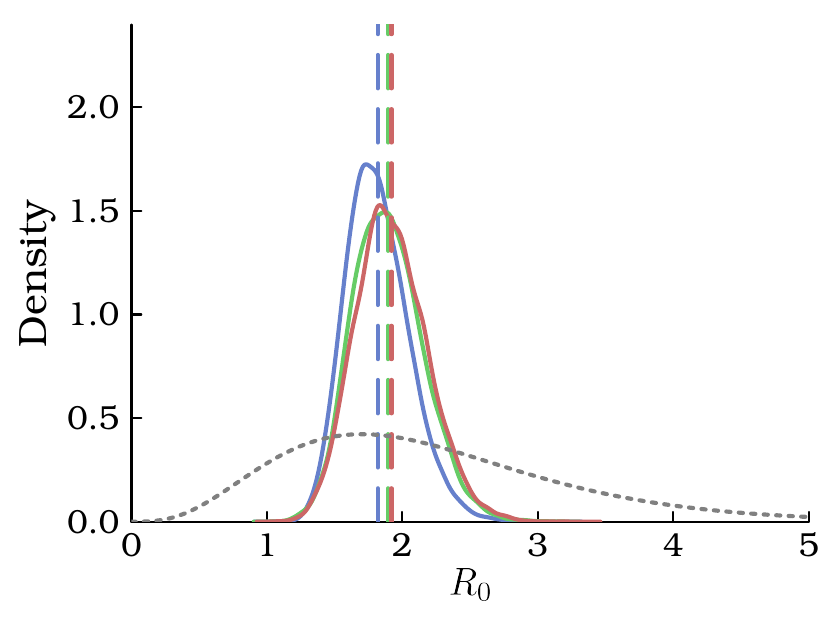}
    \end{minipage}
    \caption{\label{FIG:Vary S and p SE8I8R}(Left) Posterior distributions for three different values of the observation variance (regularisation parameter) \(\sigma^2=0.25,\,1,\,4\), (top, middle and bottom, respectively) using the three methods to evaluate the likelihood for the SE8I8R model. (Right) Posterior distributions for three different values of the observation probability \(p=0.5,\,0.75,\,1\), (top, middle and bottom, respectively) using the three methods to evaluate the likelihood for the SE8I8R model. The vertical lines are at the mean of the posterior distributions.}
\end{figure}

We fix \(T=25\) and \(R_0=2.8\) and vary the observation variance \(\sigma^2=0.25,1,4\). Figure~\ref{FIG:Vary S and p SE8I8R}~(left) shows similar results to Figure~\ref{FIG:Vary S and p SEIR}~(left). Table~\ref{TAB:Ex2 Vary S SEIR} shows the effective sample size and run time as \(\sigma^2\) varies. Table~\ref{TAB:Ex2 Vary S SEIR} demonstrates that for the smallest value of \(\sigma^2\) the run time of the particle filter method increases. Furthermore, the ESS of the PF also decreases as \(\sigma^2\) decreases; comparing Table~\ref{TAB:Ex2 Vary S SEIR} and Table~\ref{TAB:Vary S SEIR}, this decrease in ESS for the PF is more pronounced for the SE8I8R model than the SEIR model. For the Gaussian approximation the effective sample size and run time do not appear to depend on \(\sigma^2\)
\begin{table}
    \centering 
    \begin{tabular}{ccrrr}$\sigma^2$ & Method & ESS & ESS/sec & Time (s)\\ \hline \hline 
0.25 & Gaussian & 13200.9 & 6.2 & 2118.7 \\ 
0.25 & Hybrid & 3730.3 & 1.3 & 2899.3 \\ 
0.25 & Particle & 2872.2 & 0.7 & 3863.6 \\ \hline 
1.0 & Gaussian & 12762.3 & 5.7 & 2245.3 \\ 
1.0 & Hybrid & 8813.5 & 3.0 & 2976.7 \\ 
1.0 & Particle & 7136.7 & 1.9 & 3795.4 \\ \hline 
4.0 & Gaussian & 12798.0 & 7.3 & 1743.7 \\ 
4.0 & Hybrid & 11198.7 & 5.0 & 2243.9 \\ 
4.0 & Particle & 10619.6 & 3.8 & 2782.4 \\ \hline 

    \end{tabular}
    \caption{Table of effective sample sizes and run times for three different values of the observation variance (regularisation parameter) \(\sigma^2=0.25,\,1,\,4\), (top, middle and bottom, respectively) using the three methods to evaluate the likelihood for the SE8I8R model. Run time, ESS and ESS/sec are calculated excluding burn in.}
    \label{TAB:Ex2 Vary S SEIR}
\end{table}

\subsubsection*{Changing the observation probability}
We fix \(T=25\) and \(R_0=2.8\) and vary the observation probability \(p=0.5,0.75,1\). Figure~\ref{FIG:Vary S and p SE8I8R}~(right) shows the posterior distributions for this case. Similar to the SEIR model, we estimate lower values of \(R_0\) with higher \(p\). Figure~\ref{FIG:Vary S and p SE8I8R}~(right) shows no strong evidence that the accuracy of the approximation is affected by \(p\). Table~\ref{TAB:Ex2 Vary p SEIR} shows no significant effect of \(p\) on ESS for the Gaussian and PF methods; the ESS of the Hybrid method decreases as \(p\) gets larger, which is probably caused by the lower agent counts in this case compared to the others which means the switching threshold is not crossed as often. The run time of the particle filter is most affected by \(p\) and is largest for small values of \(p\), consistent with larger agent counts in the latent states of the branching process. The run time of the Gaussian approximation is unaffected by \(p\).

\begin{table}
    \centering 
    \begin{tabular}{ccrrr}$p$ & Method & ESS & ESS/sec & Time (s)\\ \hline \hline 
0.5 & Gaussian & 12731.6 & 6.1 & 2077.0 \\ 
0.5 & Hybrid & 10220.0 & 3.7 & 2749.6 \\ 
0.5 & Particle & 7366.5 & 1.3 & 5595.0 \\ \hline 
0.75 & Gaussian & 12762.3 & 5.7 & 2245.3 \\ 
0.75 & Hybrid & 8813.5 & 3.0 & 2976.7 \\ 
0.75 & Particle & 7136.7 & 1.9 & 3795.4 \\ \hline 
1.0 & Gaussian & 13088.3 & 6.3 & 2089.4 \\ 
1.0 & Hybrid & 6384.5 & 2.3 & 2729.1 \\ 
1.0 & Particle & 6252.2 & 2.5 & 2454.0 \\ \hline 

    \end{tabular}
    \caption{Table of effective sample sizes and run times for three different values of the observation probability \(p=0.5,\,0.75,\,1\), (top, middle and bottom, respectively) using the three methods to evaluate the likelihood for the SE8I8R model. Run time, ESS and ESS/sec are calculated excluding burn in.}
    \label{TAB:Ex2 Vary p SEIR}
\end{table}

\clearpage
\section{Branching Process Mean and Variance Calculation}
\label{APP::Branching Process Mean and Variance Calculation}
\cite{dormanGardenBranchingProcesses2004} gives expressions for the mean and variance of a CTBP which we now recount in the context of this paper. \cite{dormanGardenBranchingProcesses2004} considers general immigration rates (i.e., time-dependent). Since we consider constant (time-homogeneous) immigration rates only, then we can simplify the expressions for the mean and variance of the CTBP slightly. To do so, consider an augmented process \(\{\widetilde{\mathbf{z}}_t\}_{t\geq 0}\) with \(r+1\) types, the first \(r\) of which are those of the original CTBP \(\{\mathbf{z}_t\}_{t\geq 0}\) and the last of which is used to capture immigration. Hence, we have \(\widetilde{z}_{i,t} = z_{i,t}\) for \(i=1,...,r,\) and \(\widetilde{z}_{r+1,t}\) is an artificial state which captures immigration. For individuals of type \(r+1\) deaths occur at a rate equal to the net rate of immigration, \(\omega_{r+1}=\sum_{i=1}^{r}\alpha_i\) and, upon a death, the agent replaces itself and produces an offspring of type \(i=1,...,r\), with probability \(\alpha_i/\omega_{r+1}\). Further, we suppose that there is initially one agent of type \(r+1\) (and therefore there is only ever one agent of type \(r+1\)). We claim that the first \(r\) elements of the process \(\{\widetilde{\mathbf{z}}_t\}_{t\geq 0}\) are equal to the process \(\{\mathbf{z}_t\}_{t\geq 0}\) with constant immigration rates of individuals of type \(i\) equal to \(\alpha_i,\,i=1,...,r\).

In the following we show expressions for general CTBPs without explicitly capturing external immigration. From the discussion above the expressions must also apply to CTBPs with constant immigration rates as we can apply the expressions to the augmented process.



\subsubsection*{Mean Calculation}

From \cite{dormanGardenBranchingProcesses2004} the expected value of the CTBP at time \(t+s\) given the value of the process at time \(t\) is 
\[
\E{{\textbf{z}}_{t+s}|{\textbf{z}}_t}=  {\textbf{z}}_t e^{{\boldsymbol{\Omega}}s},
\]
where the exponential is the matrix-exponential. For our application, the time-step is always \(s=1\), so for convenience we define the matrix \(\textbf{F} = e^{{\boldsymbol{\Omega}}}.\)

\subsubsection*{Variance Calculation}
Consider the variance $\textbf{V}_{i,t}$ of the CTBP at time \(t\) given there is a single agent of type $i$ at time $0$,
\begin{equation}
    \label{EQN::Varince matrix}
    \textbf{V}_{i,t} := \text{Var}(\textbf{z}_{t}|\textbf{z}_0 = \textbf{u}_i).
\end{equation}
First, we need some notation. Define the matrices \(\textbf{C}_i=\omega_i\textbf{G}_i\) with 
\[
    \left[\textbf{G}_i\right]_{k,\ell} = \left[\sum_{\textbf{j}\in (\mathbb Z^+)^{r+1}} (\textbf{j}-\textbf{u}_i)\Tr(\textbf{j}-\textbf{u}_i) p_{i,\textbf{j}}\right]_{k,\ell},
\]
for \(i=1,...,r,\) where \(\textbf{u}_i\) are row-vectors with 1 in the \(i\)th position and 0 elsewhere. The elements of the matrix \(\textbf{G}_i\) are the expected product of the number of progeny of types \(k\) and \(\ell\) produced on the event of a death of an agent of type \(i\). Define the \(\text{Vec}[\cdot]\) operator which when applied to a matrix results in a column vector containing the columns of the matrix stacked on top of each other. Hence, \(\text{Vec}[\textbf{V}_{i,t}]\) (\(\text{Vec}[\textbf{C}_i]\)) is the column-vector obtained by stacking the columns of \(\textbf{V}_{i,t}\) (respectively, \(\textbf{C}_i\)) on top of each other. It is convenient to also define \(\textbf{V}_t\) (\(\textbf{C}\)) as the \(r^2\times r\) matrix obtained by stacking the vectors \(\text{Vec}[\textbf{V}_{i,t}]\) as columns of \(\textbf{V}_t\) (\(\textbf{C}\)). The vector $\text{Vec}[\textbf{V}_t]$ ($\text{Vec}[\textbf{C}]$) is the column vector obtained by stacking the columns of $\textbf{V}_t$ (respectively, $\textbf{C}$) on top of each other. Using this notation then \citep[Eq.~A3]{dormanGardenBranchingProcesses2004}, 
\begin{equation}
    \label{EQN::Integral variance}
    \text{Vec}[\textbf{V}_t]
    =
    \int_0^t
    e^{\tau{\boldsymbol{\Omega}}}
    \otimes
    e^{(t - \tau){\boldsymbol{\Omega}}\Tr}
    \otimes
    e^{(t - \tau){\boldsymbol{\Omega}}\Tr}
    \text{Vec}[\textbf{C}]\d{\tau},
\end{equation}
where $\otimes$ denotes the Kronecker product. The integrand in Equation \eqref{EQN::Integral variance} can be rewritten using the identity \citep{dormanGardenBranchingProcesses2004}
\[
(\textbf{S}\Tr\otimes \textbf{Q})\text{Vec}[\textbf{R}] = \text{Vec}[\textbf{QRS}],
\]
where \(\textbf{Q},\, \textbf{R},\) and \(\textbf{S}\) are matrices of compatible dimensions so that the left and right-hand sides are well-defined. Let \(\textbf{S}=e^{\tau{\boldsymbol{\Omega}}\Tr},\, \textbf{Q}=e^{(t - \tau){\boldsymbol{\Omega}}\Tr} \otimes e^{(t - \tau){\boldsymbol{\Omega}}\Tr}=e^{(t - \tau)\left({\boldsymbol{\Omega}}\Tr\oplus{\boldsymbol{\Omega}}\Tr\right)}\) and \(\textbf{R}=\textbf{C},\) then \((\textbf{S}\Tr\otimes \textbf{Q})\text{Vec}[\textbf{R}]\) is the integrand in \eqref{EQN::Integral variance}. Applying the identity then the integrand in \eqref{EQN::Integral variance} can be written as 
\[
\text{Vec}\left[e^{(t - \tau)\left({\boldsymbol{\Omega}}\Tr\oplus{\boldsymbol{\Omega}}\Tr\right)}\textbf{C}e^{\tau{\boldsymbol{\Omega}}\Tr}\right].
\] 
Inverting the \(\text{Vec}\) operator, then we can write 
\begin{equation}
    \label{EQN::Integral variance 2}
    \textbf{V}_t
    =
    \int_0^t
    e^{(t - \tau)\left({\boldsymbol{\Omega}}\Tr\oplus{\boldsymbol{\Omega}}\Tr\right)}\textbf{C}e^{\tau{\boldsymbol{\Omega}}\Tr}\d{\tau}.
\end{equation}

To compute \eqref{EQN::Integral variance 2} we use the identity 
\begin{eqnarray*}
    \exp\left(\begin{bmatrix} \textbf{T} & \textbf{R} \\ \textbf{0} & \textbf{U}
    \end{bmatrix}t\right) 
    &= \begin{bmatrix} e^{\textbf{T}t} & \int_{\tau=0}^te^{\textbf{T}(t-\tau)}\textbf{R} e^{\textbf{U}\tau}\d{\tau}\\ \textbf{0} & e^{\textbf{U}t}
    \end{bmatrix}
\end{eqnarray*}
with \(\textbf{T} = {\boldsymbol{\Omega}}\Tr\oplus{\boldsymbol{\Omega}}\Tr,\, \textbf{U}={\boldsymbol{\Omega}}\Tr\) and \(\textbf{R}\) as before, which gives 
\begin{eqnarray*}
    \exp\left(\begin{bmatrix} {\boldsymbol{\Omega}}\Tr\oplus{\boldsymbol{\Omega}}\Tr & \textbf{C} \\ \textbf{0} & {\boldsymbol{\Omega}}\Tr
    \end{bmatrix}t\right) 
    &= \begin{bmatrix} e^{\left({\boldsymbol{\Omega}}\Tr\oplus{\boldsymbol{\Omega}}\Tr\right)t} & \int_{\tau=0}^te^{\left({\boldsymbol{\Omega}}\Tr\oplus{\boldsymbol{\Omega}}\Tr\right)(t-\tau)}\textbf{C} e^{{\boldsymbol{\Omega}}\Tr\tau}\d{\tau}\\ \textbf{0} & e^{{\boldsymbol{\Omega}}\Tr t}
    \end{bmatrix}.
\end{eqnarray*}
Notice that the expression in \eqref{EQN::Integral variance 2} is the upper-right quadrant of size \(r^2\times r\). Additionally, the lower-right quadrant of size \(r \times r\) is the transpose of the mean operator \(e^{\boldsymbol{\Omega}\Tr t}\).

By the independence of agents the variance for an arbitrary starting state \(\textbf{z}_0\) is 
\[
    \text{Var}(\textbf{z}_{t}|\textbf{z}_0 ) = \sum_{i=1}^r \textbf{V}_{i,t}z_{0,i},
\]
and, by the time-homogeneous property of CTBPs, for any \(t,s\geq 0\),
\[
    \text{Var}(\textbf{z}_{t+s}|\textbf{z}_t) = \sum_{i=1}^r \textbf{V}_{i,s}z_{t,i}.
\]
In fact, 
\begin{align}
    \begin{bmatrix}
        \text{Vec}\left[\text{Var}(\textbf{z}_{t+s}|\textbf{z}_t )\right] \\
        \text{Vec}\left[\E{{\textbf{z}}_{t+s}|{\textbf{z}}_t}\right]
    \end{bmatrix} 
    = \exp\left(\begin{bmatrix} {\boldsymbol{\Omega}}\Tr\oplus{\boldsymbol{\Omega}}\Tr & \textbf{C} \\ \textbf{0} & {\boldsymbol{\Omega}}\Tr
    \end{bmatrix}s\right) 
    \begin{bmatrix}
        \boldsymbol{0} \\
        {\textbf{z}}_t\Tr
    \end{bmatrix}. \label{EQN:: var matrix exp}
\end{align}

Since this work uses time steps of \(s=1\) only, then it is useful to define the notation \(\textbf{V}_i:=\textbf{V}_i(1)\).

\section{Kalman Filter Approximation for CTBPs}
\label{SEC::Filtering Distribution}
Here we provide details on how to apply the Kalman filter approximation to CTBPs. Let \(\{\textbf{z}_t\}_{t\geq0}\) be a CTBP and consider the observation process in Equation \eqref{EQN::Obsevation dist requirement}. Define \(\textbf{x} \mapsto\phi(\textbf{x}; \boldsymbol{\mu}, \boldsymbol{\Sigma})\) as the density function of a multivariate normal distribution with mean \(\boldsymbol{\mu}\) and covariance \(\boldsymbol{\Sigma}\). With this notation, given \(\textbf{z}_t\), the density of the observation process is \(\phi(\textbf{y}_t; \textbf{H}\textbf{z}_t\Tr, \textbf{R})\).

The starting point of the filtering iterations is the posterior distribution \(p(\textbf{z}_t\cond\textbf{y}_{1:t})\), from which we obtain the mean \(\boldsymbol{\mu}_{t\cond t}:=\E{\textbf{z}_t\cond \textbf{y}_{1:t}}\) and variance \(\boldsymbol{\Sigma}_{t\cond t}:=\Var{\textbf{z}_t\cond \textbf{y}_{1:t}}\). As discussed in Section~\ref{SEC::Transition Density}, the mean and variance are then propagated to time \(t+1\) using the CTBP, that is, 
\begin{align}
    \label{EQN::KF 1}
    \boldsymbol{\mu}_{t+1\cond t} &= \boldsymbol{\mu}_{t\cond t}\textbf{F}
    \\\label{EQN::KF 2}
    \boldsymbol{\Sigma}_{t+1\cond t} &= \sum_{i=1}^r\mu_{i,t\cond t} \textbf{V}_i + \textbf{F}\Tr\boldsymbol{\Sigma}_{t\cond t}\textbf{F},
\end{align}
where \(\mu_{i,t\cond t}\) is the \(i\)th element of \(\boldsymbol{\mu}_{t\cond t}\). This is equivalent to applying the prediction step of a Kalman filter with movement model \(\textbf{F}\) and movement noise covariance \(\sum_{i=1}^r\mu_{i,t\cond t} \textbf{V}_i\).

The distribution \(p(\textbf{z}_{t+1}\cond\textbf{y}_{1:t})\) is then approximated as a Gaussian distribution using \(\boldsymbol{\mu}_{t+1\cond t}\) and \(\boldsymbol{\Sigma}_{t+1\cond t}\) as the mean and variance of the distribution, i.e., 
\[
    p(\textbf{z}_{t+1}\cond\textbf{y}_{1:t}) \approx \phi(\textbf{z}_{t+1}; \boldsymbol{\mu}_{t+1\cond t}\Tr, \boldsymbol{\Sigma}_{t+1\cond t}).
\]
Now we apply the usual update step of the Kalman filter. Specifically, by Bayes Theorem
\begin{align*}
    p(\textbf{z}_{t+1}\cond\textbf{y}_{1:t+1}) 
    &\propto p(\textbf{y}_{t+1}\cond\textbf{z}_{t+1}, \textbf{y}_{1:t})p(\textbf{z}_{t+1}\cond\textbf{y}_{1:t})
    \\&\approx \phi(\textbf{y}_{t+1}; \textbf{H}\textbf{z}_{t+1}\Tr, \textbf{R})\phi(\textbf{z}_{t+1}; \boldsymbol{\mu}_{t+1\cond t}\Tr, \boldsymbol{\Sigma}_{t+1\cond t}).
\end{align*}
Therefore, the distribution \(p(\textbf{z}_{t+1}\cond\textbf{y}_{1:t+1})\) is approximately a Gaussian distribution with mean \(\boldsymbol{\mu}_{t+1\cond t+1}\Tr\) and covariance \(\boldsymbol{\Sigma}_{t+1\cond t+1}\), which can be calculated using the update step of the Kalman filter as follows.
\begin{align}
    \label{EQN::KF 3}
    \textbf{K}_{t+1} &:= \boldsymbol{\Sigma}_{t+1\cond t}\textbf{H}\Tr\left(\textbf{H}\boldsymbol{\Sigma}_{t+1\cond t}\textbf{H}\Tr+\textbf{R}\right)^{-1},
    \\\label{EQN::KF 4}
    \boldsymbol{\mu}_{t+1\cond t+1}\Tr &= \boldsymbol{\mu}_{t+1 \cond t}\Tr + \textbf{K}_{t+1}\left(\textbf{y}_{t+1}\Tr-\textbf{H}\boldsymbol{\mu}_{t+1\cond t}\Tr\right),
    \\\label{EQN::KF 5}
    \boldsymbol{\Sigma}_{t+1\cond t+1} &= (\textbf{I}-\textbf{K}_{t+1}\textbf{H})\boldsymbol{\Sigma}_{t+1\cond t}.
\end{align}
Thus, the filtering iterations amount to iterating \eqref{EQN::KF 1}-\eqref{EQN::KF 2} and \eqref{EQN::KF 3}-\eqref{EQN::KF 5}, which is a standard Kalman filter where the variance of the movement model changes at each iteration.

We use the Gaussian approximation to approximate the likelihood terms as 
\[
    p(\textbf{y}_{t+1} \cond \textbf{y}_{1:t}) \approx \phi(\textbf{y}_{t+1}; \textbf{H}\boldsymbol{\mu}_{t+1\cond t}\Tr, \textbf{H}\boldsymbol{\Sigma}_{t+1\cond t}\textbf{H}\Tr+\textbf{R}).
\]
Occasionally the state estimate, \(\boldsymbol{\mu}_{t \cond t}\), may contain elements which are negative. However, the states of a CTBP are strictly non-negative. In our experience, this is most likely to occur when agent counts are small and decreasing. When the state estimate contains negative elements we return a likelihood value of 0 and cease the Kalman filter iteration.

\bibliographystyle{elsarticle-harv}
\bibliography{overleaf_ref}

\end{document}